\documentstyle[preprint,prl,aps]{revtex}
\begin{document}
\def\sn2{$\sin^22\theta$}
\def\dm2{$\Delta m^2$}
\def\ch2{$\chi^2$}
\def\ltap{\ \raisebox{-.4ex}{\rlap{$\sim$}} \raisebox{.4ex}{$<$}\ }
\def\gtap{\ \raisebox{-.4ex}{\rlap{$\sim$}} \raisebox{.4ex}{$>$}\ }
\draft
\begin{titlepage}
\preprint{\vbox{\baselineskip 10pt{
\hbox{Ref. SISSA 174/96/EP}
\hbox{hep -- ph/9702400}
\hbox{February 1997}}}}
\vskip -0.4cm
\title{ \bf Three-Neutrino
Mixing and Combined Vacuum Oscillations and MSW Transitions of Solar Neutrinos 
          }
\author{Q.Y. Liu $^{a)}$ and S.T. Petcov $^{a,b)}$\footnote{Also at:
Institute of Nuclear Research and Nuclear Energy, Bulgarian Academy
of Sciences, BG--1784 Sofia, Bulgaria.}}
\address{a) Scuola Internazionale Superiore di Studi Avanzati, I-34013
Trieste, Italy}
\address{b) Istituto Nazionale di Fizica Nucleare, Sezione di Trieste, I-34013 
Trieste, Italy}  
\maketitle
\begin{abstract}
\begin{minipage}{5in}
\baselineskip 16pt
Assuming three flavour neutrino mixing takes place in vacuum, 
we investigate the possibility
that the solar $\nu_e$ take part in MSW transitions in the Sun
due to $\Delta m^2_{31} \sim (10^{-7} - 10^{-4})~eV^2$, followed by long
wave length vacuum oscillations on the way to the Earth, triggered by
$\Delta m^2_{21}$ (or $\Delta m^2_{32}$) $\sim (10^{-12} - 10^{-10})~eV^2$,
$\Delta m^2_{31}$ and $\Delta m^2_{21}$ ($\Delta m^2_{32}$) being the corresponding
neutrino mass squared differences.
The solar $\nu_e$ survival probability is shown to be described in this 
case by a simple analytic expression. Depending on whether 
the vacuum oscillations are due to 
$\Delta m^2_{21}$ or $\Delta m^2_{32}$ there 
are two very different types of interplay between 
the MSW transitions and the vacuum oscillations of the solar $\nu_e$. 
Performing an analysis of the most recently published solar neutrino data
we have found several qualitatively 
new solutions of the solar neutrino problem  of the hybrid
MSW transitions + vacuum oscillations type. The solutions differ
in the way the $pp$, $^{7}$Be and $^{8}$B neutrino fluxes are
affected by the transitions in the Sun and the oscillations in vacuum.
The specific features of the new solutions are discussed.  
\end{minipage}
\end{abstract}
\end{titlepage}
\newpage

\hsize 16.5truecm
\vsize 24.0truecm
\def\dm{$\Delta m^2$\hskip 0.1cm }
\def\dmsqua{$\Delta m^2$\hskip 0.1cm}
\def\sn{$\sin^2 2\theta$\hskip 0.1cm }
\def\snf{$\sin^2 2\theta$}
\def\trna{$\nu_e \rightarrow \nu_a$}
\def\trnm{$\nu_e \rightarrow \nu_{\mu}$}
\def\trns{$\nu_e \leftrightarrow \nu_s$}
\def\trnat{$\nu_e \leftrightarrow \nu_a$}
\def\trnmt{$\nu_e \leftrightarrow \nu_{\mu}$}
\def\trne{$\nu_e \rightarrow \nu_e$}
\def\trnst{$\nu_e \leftrightarrow \nu_s$}
\def\nue{$\nu_e$\hskip 0.1cm}
\def\numu{$\nu_{\mu}$\hskip 0.1cm}
\def\nutau{$\nu_{\tau}$\hskip 0.1cm}

\font\eightrm=cmr8
\def\aprle{\buildrel < \over {_{\sim}}}
\def\aprge{\buildrel > \over {_{\sim}}}
\renewcommand{\thefootnote}{\arabic{footnote}}
\setcounter{footnote}{0}
\leftline{\bf 1. Introduction}
\vskip 0.3cm

\indent It is well known that the solar neutrino problem 
\cite{CHLOR,KAM,GALLEX,SAGE,SNP,SNP1,SNP2}
admits, in particular, two quite different neutrino physics 
solutions: one based on the 
old idea of Pontecorvo \cite{Pont1} that 
the solar $\nu_e$ take part in vacuum oscillations
when they travel from the Sun to the Earth, and a second based on the more 
recent hypothesis \cite{MS,LW} of the solar $\nu_e$ undergoing matter-enhanced (MSW)
transitions into neutrinos of a different type when they propagate 
from the central part to the surface of the Sun. Both these solutions and
the possibilities
to test them and to distinguish between them in the future solar neutrino 
experiments, have been extensively studied in the simplest case of 
two-neutrino mixing (see, e.g., refs. \cite{KP1,VO,KP2,MSW1,MSW2,KPnu96} and the
articles quoted therein). The vacuum oscillations and 
the MSW transitions of the
solar \nue are characterized by the same two parameters 
in this case \cite{Pont1,MS,LW,SMBP78}:
\snf, $\theta$ being the neutrino (lepton) mixing angle in vacuum, and
\dmsqua = $m^2_2$ - $m^2_1 > 0$, where $m_{1,2}$ are the masses of two neutrinos 
$\nu_{1,2}$ with definite mass in vacuum. The vacuum oscillation interpretation
of the solar neutrino problem requires that \cite{KPnu96}
$$5.0 \times 10^{-11}eV^2 \ltap \Delta m^2 
      \ltap 1.1\times 10^{-10}eV^2,~~\eqno(1a)$$
  $$0.67 \ltap sin^22\theta \leq 1.0, ~~\eqno(1b)$$
while the MSW solution is possible for \cite{KPnu96}
  $$3.6 \times 10^{-6}eV^2 \ltap \Delta m^2 
        \ltap 9.8\times 10^{-6}eV^2,~~\eqno(2a)$$
  $$4.5\times 10^{-3} \ltap sin^22\theta \ltap 1.3\times 10^{-2}, ~~\eqno(2b)$$
and for 
  $$5.7 \times 10^{-6}eV^2 \ltap \Delta m^2 
            \ltap 9.5\times 10^{-5}eV^2,~~\eqno(3a)$$
  $$0.51 \ltap sin^22\theta \ltap 0.92, ~~\eqno(3b)$$

\noindent if the MSW transitions are into an active neutrino, $\nu_{\mu (\tau)}$   
\footnote{For a detailed study of the MSW solution of the solar neutrino problem
with solar \nue transitions into a sterile neutrino, $\nu_s$, see ref. \cite{KLP}.}.
The \dmsqua solution interval (1a) corresponds to values of
the solar \nue oscillation length in vacuum comparable to the Sun-Earth distance.

   Detailed studies of the vacuum oscillation or MSW
transition solution of the solar neutrino problem
under the more natural assumption of three
flavour neutrino mixing are still lacking, partly because
of the relatively large number of parameters involved. The three 
flavour neutrino mixing hypothesis implies the existence of three neutrinos
$\nu_k$ with definite vacuum masses, $m_k$, $k=1,2,3$, and the relevant
vacuum oscillation and/or MSW transition probabilities depend now on 
two different neutrino mass squared differences
\footnote{One can always choose   
$m_1 < m_2 <m_3$ without loss of generality 
and we will work with this convention in what follows.}, $\Delta m^2_{21} > 0$ 
and $\Delta m^2_{31} > 0$, 
and on at least two
mixing angles, say, $\theta_{12}$ and $\theta_{13}$. 
The general features of the three-neutrino MSW transitions 
of solar neutrinos have been analyzed
by many authors (see, e.g., refs. \cite{3nuMSW,3nuSP}). 
Certain aspects of the 
possible three-neutrino MSW solution of the solar neutrino problem 
have been discussed as well \cite{MS88,AS91}. However, comprehensive results
based on the more recent and more precise data from the four solar
neutrino experiments have been obtained only  
in specific cases when, for instance, 
$\Delta m^2_{21}$ has a value 
in one of the solution intervals (2a), (3a), 
while $\Delta m^2_{31}>> 10^{-4}~eV^2$ \cite{SHI,Lisi}. Under these 
conditions the solar \nue MSW transitions are essentially
of the two-neutrino mixing type \cite{3nuMSW,3nuSP}, although the regions of 
solutions can differ somewhat from those given in eqs. (2a) -- (3b) \cite{SHI,Lisi}.

\indent In the present article we study the qualitatively new possibility of a
hybrid vacuum oscillation and MSW transition solution of the 
solar neutrino problem {\footnote{An earlier rather brief and
qualitative discussion of combined solar $\nu_e$ MSW transitions and long wave length
vacuum oscillations in connection with the solar
neutrino problem can be found in ref. \cite{AS91}.}. 
This possibility is rather natural if 
three-neutrino mixing takes place in vacuum.

 One would guess a priori that such a solution would require $\Delta m^2_{21}$
(or $\Delta m^2_{32}$) and $\Delta m^2_{31}$ to have values 
in the regions specified by eq. (1a) and by eqs. (2a) or (3a), 
respectively. This corresponds to the case 
$$\Delta m^2_{21} \ll \Delta m^2_{31} \cong \Delta m^2_{{\rm 32}},~~\eqno(4)$$
or to
$$\Delta m^2_{32} \ll \Delta m^2_{31} \cong \Delta m^2_{{\rm 21}}.~~\eqno(5)$$

\indent Several patterns of neutrino masses can lead to the inequality (4) or
(5) and to the requisite values of $\Delta m^2_{21}$ (or $\Delta m^2_{32}$) and
$\Delta m^2_{31}$. Eqs. (1a), (2a) or (3a) and (4) (or (5)) 
are compatible, for instance, 
i) with the existence of three quasi-degenerate neutrinos, 
$m_1 \cong m_2 \cong m_3$, with a very different splitting between their
masses, $(m_2 - m_1) \ll  (m_3 - m_1)$,
ii) with a neutrino mass spectrum consisting of two close masses which are very 
different from the third one, ${m_1 \cong m_2 \ll m_3}$, and 
iii) with a hierarchical neutrino mass spectrum, 
$m_1 \ll m_2 \ll m_3$.  
In the first case neutrinos can 
have masses in the cosmologically
relevant region, $m_i \sim (1-2)~eV$, with \footnote{If the massive neutrinos $\nu_i$
are Majorana particles, the neutrinoless double beta 
decay would be allowed and, for $m_i \sim(1 - 2)~eV$, can proceed with a rate
which is in the range of sensitivity of the new generation 
of experiments searching for 
this decay (see, e.g., \cite{BBN0}).}
$(m_3 - m_1) \sim (10^{-6}-10^{-5})~eV$  and 
$(m_2 - m_1) \sim (10^{-12}-10^{-10})~eV$, 
while in the other two cases the neutrino masses are too small 
to be of significance for the solution of the dark matter 
problem: $m_3 \sim 10^{-3}~eV$ and  
$m_2 \sim m_1 \sim (m_2 - m_1) \sim (10^{-6}-10^{-5})~eV$ (case ii) or
$m_2 \sim (10^{-5}-10^{-6})~eV$ (case iii). In case ii) the 
pattern ${m_1 \ll m_2 \cong m_3}$ with $m_{2,3} \sim 10^{-3}~eV$ and 
$(m_3 - m_2) \sim (10^{-8}-10^{-7})~eV$ is also possible.
It corresponds to eq. (5) and arises in the Zee model of neutrino mass generation 
\cite{Zee,SPZee,ASTZee} and its various GUT generalizations. 
Obviously, if $\Delta m^2_{21}$ ($\Delta m^2_{32}$) and 
$\Delta m^2_{31}$ lie in the
intervals (1a) and (2a) or (3a) respectively, there 
will be no observable neutrino oscillation effects due to 
$\Delta m^2_{21}$ ($\Delta m^2_{32}$) and 
$\Delta m^2_{31}$ in the oscillation experiments performed with 
terrestrial (reactor, accelerator, etc.) or atmospheric neutrinos.

\indent All the different 
patterns of neutrino masses mentioned above
can arise in gauge theories of electroweak interactions with 
massive neutrinos, and in particular, in GUT theories 
and in the superstring inspired models (see, e.g., \cite{GUT}).  

\indent This paper is organized as follows. Section 2 is devoted to the discussion
of the average solar $\nu_e$ survival probability, $\bar{P}(\nu_e \rightarrow \nu_e)$,
when three-neutrino mixing takes place in vacuum and the solar neutrinos
undergo MSW transitions in the Sun, which are followed by long wave length
neutrino oscillations in vacuum. It is shown, in particular, that the
probability $\bar{P}(\nu_e \rightarrow \nu_e)$ is 
described by a simple analytic expression and that inequalities (4)
and (5) imply two very different types of interplay between the 
MSW transitions and the vacuum oscillations of the solar \nue. In Section 3
the hypothesis that the solar \nue take part in MSW transitions and long wave length
vacuum oscillations is confronted with the solar neutrino data. 
Several new hybrid MSW transition + vacuum oscillation (MSW + VO) 
solutions of the solar neutrino
problem are found and their specific features
are discussed. Section 4 contains our conclusions. A derivation
of the analytic expression for the probability $\bar{P}(\nu_e \rightarrow \nu_e)$
used in the present study is given in the Appendix.

\vglue 0.4cm
\leftline{\bf 2. The Solar \nue Survival Probability}
\vskip 0.3cm
\indent It is possible to obtain a very simple 
analytic expression for the solar \nue survival probability 
in the case of interest, namely, when the solar \nue 
undergo matter-enhanced (MSW) transitions in the Sun and 
vacuum oscillations on their way from the surface of 
the Sun to the Earth. The vacuum oscillations are supposed 
to proceed for neutrinos having energy $E \sim 1~MeV$ with an 
oscillation length comparable to the 
Sun--Earth distance, $R_0 \cong 1.4966 \times 10^8~km$. 

\indent We assume three-flavour neutrino mixing takes place in vacuum, 
$$|\nu_l>~ = \sum_{k=1}^3 U_{lk}^* |\nu_k>, \hspace{1cm} l=e,\mu,\tau, ~~\eqno(6)$$ 
\noindent where $|\nu_l>$ is the state vector of 
the (left-handed) flavour neutrino $\nu_l$ having momentum 
$\overrightarrow{p}$, $|\nu_k>$ is the state vector of a neutrino 
$\nu_k$ possessing a definite mass $m_k$ and momentum $\overrightarrow{p}$, 
$m_k \neq m_j$, $k \neq j = 1,2,3$, $m_1 < m_2 <  m_3$, and U is a $3 \times 3$
unitary matrix -- the lepton mixing matrix. 
The neutrinos $\nu_k$ are assumed to be stable and relativistic.
We shall suppose also that the matter-enhanced 
transitions of the solar \nue into \numu and/or \nutau in the Sun are 
associated with $\Delta m^2_{31}$, where, as usual,  
$\Delta m^2_{ij}= m^2_i-m^2_j$, while $\Delta m^2_{21}$ 
(or $\Delta m^2_{32}$) is responsible for the long wave length 
($\sim 1.5\times 10^8~km$) $\nu_e \leftrightarrow \nu_{\mu,\tau}$
oscillations taking place between the Sun and the Earth. 
This implies that either relation (4) or (5) is valid . We shall 
consider in the present analysis a somewhat wider range 
of values of $\Delta m^2_{21}$ ($\Delta m^2_{32}$) and 
$\Delta m^2_{31}$ than is given by the two-neutrino mixing 
solution intervals eqs. (1a) and (2a) or (3a), respectively: 
     $$10^{-12}~eV^2 \leq \Delta m^2_{21} \ltap 5.0\times 10^{-10}~eV^2,~~\eqno(7)$$
     $$10^{-7}~eV^2 \leq \Delta m^2_{31} \leq 10^{-4}~eV^2.~~\eqno(8)$$

  Under the above conditions and assuming the inequality (4),  
$\Delta m^2_{21} \ll \Delta m^2_{31}$, holds, 
the analytic expression for the solar
\nue survival probability of 
interest, $\bar{P}(\nu_e \rightarrow \nu_e)$, can be easily
deduced from the general expression for $\bar{P}(\nu_e \rightarrow \nu_e)$ derived
in ref. \cite{3nuSP} (see eq. (29) in \cite{3nuSP}) for the relevant case of 
three (flavour) neutrino mixing. It has the following simple form:
$$\bar{P}(\nu_e \rightarrow \nu_e ; t_E,t_0) = 
 \bar{P}^{(31)}_{2MSW}(\nu_e \rightarrow \nu_e ;t_{\odot},t_0)~~~~~~~~~~~~~~~~~~~
~~~~~~~~~~~~~~~~~~~~~~~~~~~~~~~~~~~~~~~~~$$ 
$$~~~~~~~~~~~~~~~~~ +~~{{1 - |U_{e3}|^2}\over {1 - 2|U_{e3}|^2}}~[1 -  
  P^{(21)}_{2VO}(\nu_e \rightarrow \nu_e ;t_E,t_{\odot})]~
  [|U_{e3}|^2 - \bar{P}^{(31)}_{2MSW}(\nu_e \rightarrow \nu_e ;t_{\odot},t_0)].
         ~\eqno(9)$$
\vskip 0.2truecm
\noindent Here $\bar{P}(\nu_e \rightarrow \nu_e ;t_E,t_0)$ is the average 
probability that a solar 
\nue having energy $E \simeq |\overrightarrow{p}| = p$ will not be converted 
into \numu and/or \nutau 
when it propagates from the central part of the Sun, 
where it was produced at time $t_0$, to the surface of 
the Earth reached at time $t_E$, 
$\bar{P}^{(31)}_{2MSW}(\nu_e \rightarrow \nu_e ;t_{\odot},t_0)$ 
is the average solar
\nue survival probability when the \nue undergoes two-neutrino
MSW transitions due to $\Delta m^2_{31}$ and $U_{e3} \neq 0$ on its way
to the surface of the Sun, $t_{\odot}$ being the time at which
the \nue reaches the surface of the Sun, and 
$P^{(21)}_{2VO}(\nu_e \rightarrow \nu_e ;t_E,t_{\odot})$ is the solar \nue survival 
probability for a \nue taking part in two-neutrino vacuum oscillations
due to $\Delta m^2_{21}$ and $U_{e1} \neq 0$, $U_{e2} \neq 0$, 
on its way from the surface of the Sun to the surface of the Earth.
The two-neutrino MSW transition and vacuum oscillation probabilities 
$\bar{P}^{(31)}_{2MSW}(\nu_e \rightarrow \nu_e ;t_{\odot},t_0)$
and $P^{(21)}_{2VO}(\nu_e \rightarrow \nu_e ;t_E,t_{\odot})$
are given by the well-known expressions \cite{Pont1,MS,SMBP78,Haxt87,SP191,SP200,BPet87}:
$$\bar{P}^{(31)}_{2MSW}(\nu_e \rightarrow \nu_e ;t_{\odot},t_0)~\equiv 
~\bar{P}^{(31)}_{2MSW}~=~{1\over {2}} + ({1\over {2}} - P'_{(31)})
 \cos2\theta_{13} \cos2\theta_{13}^{m}(t_0),~~~\eqno(10)$$
$$P^{(21)}_{2VO}(\nu_e \rightarrow \nu_e ;t_E,t_{\odot})~\equiv~P^{(21)}_{2VO}~=
1 - {1\over {2}}~\sin^22\theta_{12}~(1 - 
              \cos2\pi {R \over {L^{v}_{21}}}).~~~\eqno(11)$$ 

\noindent In eqs. (10) and (11) $P'_{(31)}$ is the so-called ``jump'' 
(or ``level-crossing'') probability,
$$\cos2\theta_{13} = 1 - 2|U_{e3}|^2 > 0,~~
    \sin2\theta_{13} = 2|U_{e3}|\sqrt{1 - |U_{e3}|^2}~> 0,~~~~\eqno(12)$$
$\theta_{13}^{m}$ is the two-neutrino mixing angle in matter, 
which in vacuum coincides with $\theta_{13}$,
$$\cos2\theta_{13}^{m}(t_0) = 
{{1 - N_e(t_0)/N_e^{res}}\over 
{\sqrt{(1 - {N_e(t_0)\over {N_e^{res}}})^2 + \tan^22\theta_{13}}}},~~~\eqno(13)$$
where $N_e(t_0)$ is the electron number 
density at the point of $\nu_e$ production in
the Sun and $ N_e^{res} = \Delta m^2_{31}\cos2\theta_{13}/(2p\sqrt{2} G_F)$ 
is the resonance density,
$$\sin^22\theta_{12} = 4 {{|U_{e1}|^2~|U_{e2}|^2}\over {(|U_{e1}|^2 + 
|U_{e2}|^2)^2}}~~,~~~~\eqno(14)$$ 
$R = (t_E - t_{\odot})$ is the distance traveled by the neutrinos in vacuum,
$R \cong R_0$, and $L^{v}_{21} = 4\pi p/\Delta m^2_{21}$ is the oscillation length
in vacuum associated with $\Delta m^2_{21}$. 

 Several comments are in order. The requirement  
$$|U_{e3}|^2 < 0.5~~~~\eqno(15)$$ 
in eq. (12) ensures the possibility of a resonant 
enhancement of the solar $\nu_e$ transitions in the Sun \cite{ResCon,MS}.

 For the values of $\Delta m^2_{31}$ of interest, eq. (8), the oscillating terms in
the probability $\bar{P}(\nu_e \rightarrow \nu_e ; t_E,t_0)$, associated with
the solar $\nu_e$ MSW transitions in the Sun \cite{SP88osc} are rendered 
negligible \cite{SPJR89} by the various averagings one has to perform
when calculating the effects of the transitions and oscillations on the signals
in the solar neutrino detectors. 
  
   Further, the probability 
$\bar{P}(\nu_e \rightarrow \nu_e ; t_E,t_0)$, eq. (9), depends only on the 
absolute values of the elements of the lepton mixing matrix $U$, 
forming the first row of $U$, more precisely, on $|U_{ek}|^2$, $k=1,2,3$. Of these
only two are independent since the unitarity of $U$ implies: 
$|U_{e1}|^2 + |U_{e2}|^2 + |U_{e3}|^2 = 1$. It follows from eqs. (12), (14) and 
the preceding remark that $\theta_{13}$  
and $\theta_{12}$ are independent parameters.

  We note next that at least three different expressions 
for the ``jump'' probability $P'_{(31)}$ entering into formula (10) for  
the MSW probability have been proposed
\cite{Haxt87,SP200}. We will use 
the one derived in ref. \cite{SP200} in the exponential density approximation:
$$P'_{(31)} = {{e^{-2\pi r_{0}{\Delta m^2_{31}\over {2p}}\sin^2\theta_{13}}
          - e^{-2\pi r_{0}{\Delta m^2_{31}\over {2p}}}}
         \over {1 - e^{-2\pi r_{0}{\Delta m^2_{31}\over {2p}}}}}~~,~\eqno(15)$$
\vskip 0.2truecm
\noindent where $r_{0}$ is the scale-height of the variation
of the electron number density $N_e$ along the neutrino path in the Sun.
Unlike the expressions proposed in ref. \cite{Haxt87}, the one
given by eq. (15) describes correctly the (strongly) nonadiabatic MSW 
transitions of solar neutrinos for values of $\sin^22\theta_{13} \gtap (0.2 - 0.3)$ 
\cite{SP191,SP200,KP88}. 
 
  Let us comment on how eq. (9) with probabilities   
$\bar{P}^{(31)}_{2MSW}$ and $P^{(21)}_{2VO}$ given by eqs. (10) and (11) 
can be obtained from eq. (29) in ref. \cite{3nuSP}. The analysis leading to eq. (29) in
\cite{3nuSP} is valid, in particular, under the condition 
$\Delta m^2_{21} \ll \Delta m^2_{31}$. This condition is fulfilled in the case
presently considered, which is specified by eqs. (4), (7) and (8). We observe first
that in this case one has, in the notations of ref. \cite{3nuSP}, 
$\tilde{c}_{12}\cong 0$,
$\varphi'_{13}(t'_0)\cong 0$ (and therefore $\tilde{c}_{13}\cong c_{13}$,
$\tilde{s}_{13}\cong s_{13}$) and $s'_{23} \cong 0$ in eq. (29); as a consequence,
the latter simplifies considerably. We notice next that for
$\Delta m^2_{21}$ having values in the interval (7) 
the times $t$ and $t'_0$ in eq. (29)
in \cite{3nuSP} can be considered to be respectively the times $t_E$ and $t_{\odot}$ with
the meaning they have in eq. (9) above: the derivation of eq. (29) in \cite{3nuSP}
for $\bar{P}(\nu_e \rightarrow \nu_e)$
is valid in this case as well. Finally, i) the angles $\varphi_{12}$ and 
$\varphi_{13}$ in eq. (29) in \cite{3nuSP} coincide with the angles $\theta_{12}$ 
and $\theta_{13}$ defined by eqs. (12) and (14) above, 
ii) the probabilities $\bar{P}(\nu_e \rightarrow \nu_e ;t,t_0)$, 
$\bar{P}_{H}(\nu_e \rightarrow \nu_e; t'_0,t_0)$ and
$\bar{P}_{L}(\nu_e \rightarrow \nu_e; t,t'_0)$ in eq. (29) coincide for 
$t \equiv t_E$ and $t'_0 \equiv t_{\odot}$ with the probabilities
$\bar{P}(\nu_e \rightarrow \nu_e ;t_E,t_0)$,
$\bar{P}^{(31)}_{2MSW}(\nu_e \rightarrow \nu_e ;t_{\odot},t_0)$ 
and $P^{(21)}_{2VO}(\nu_e \rightarrow \nu_e ;t_E,t_{\odot})$, respectively, and
iii) the term ${\displaystyle \overline{{\rm Re}~[R_{H}(t'_0,t_0)}]}$ in eq. (29) 
in the case of interest is given, as it is not difficult to show using the 
results in refs. \cite{SP200,SP88osc,AASP92}, by the 
following expression:
$$\overline{{\rm Re}~[R_{H}(t'_0,t_0)]} = -~({1\over{2}} - P'_{(31)})
  \sin2\theta_{13}\cos2\theta_{13}^{m}(t_0).~~~~\eqno(16)$$   
\noindent Given the above observations it is easy to obtain expression (9)
for $\bar{P}(\nu_e \rightarrow \nu_e ;t_E,t_0)$ from eq. (29) in \cite{3nuSP}
\footnote{Note that there is a misprint in eq. (29) in \cite{3nuSP}:
the probability $\bar{P}_{L}(\nu_e \rightarrow \nu_e; t,t'_0)$ in the expression in
the curly brackets multiplying the term 
$\overline{{\rm Re}~[R_{H}(t'_0,t_0)]}$ in the third row
should read $\bar{P}_{L}(\nu_e \rightarrow \nu_{\mu}; t,t'_0) \equiv
1 - \bar{P}_{L}(\nu_e \rightarrow \nu_e; t,t'_0)$.}. A proof of eq. (9) which does
not rely on eq. (29) in \cite{3nuSP} and is 
based only on some of the intermediate results of
the analysis performed in \cite{3nuSP} is given in the Appendix.  

  Using eqs. (10), (11) and (12) we can rewrite the 
second term in eq. (9) in a form which 
proves convenient for our later discussion:
$$\bar{P}(\nu_e \rightarrow \nu_e ;t_E,t_0) = \bar{P}^{(31)}_{2MSW} 
- {1\over {2}}~\sin^22\theta_{12}~(1 - \cos2\pi {R \over {L^{v}_{21}}})~
\cos^2\theta_{13} [{1\over {2}} + ({1\over {2}} - P'_{(31)})
\cos2\theta_{13}^{m}(t_0)]~~\eqno(17)$$.
\vskip -0.2truecm
\indent Before proceeding further let us point out that since 
$\bar{P}^{(31)}_{2MSW} \geq |U_{e3}|^2 = \sin^2\theta_{13}$ \cite{MS}
and $|U_{e3}|^2 < 0.5$, for fixed values of the MSW transition parameters
one always has 
$$\bar{P}(\nu_e \rightarrow \nu_e ;t_E,t_0) \geq
{{|U_{e3}|^2}\over {1 - 2|U_{e3}|^2}}~[1 - |U_{e3}|^2 - \bar{P}^{(31)}_{2MSW}] =
\sin^2\theta_{13} [{1\over {2}} - ({1\over {2}} - P'_{(31)})
\cos2\theta_{13}^{m}(t_0)]~\eqno(18)$$
\noindent as it follows from eq. (9). This lower limit constrains effectively 
$\bar{P}(\nu_e \rightarrow \nu_e ;t_E,t_0)$ only for sufficiently large
values of $\sin^2\theta_{13}$. Setting 
$P^{(21)}_{2VO}(\nu_e \rightarrow \nu_e ;t_E,t_{\odot}) = 1$ in eq. (9)
we get an absolute upper limit on $\bar{P}(\nu_e \rightarrow \nu_e ;t_E,t_0)$
for given $\bar{P}^{(31)}_{2MSW}$ and $|U_{e3}|^2$:
$$\bar{P}(\nu_e \rightarrow \nu_e ;t_E,t_0) \leq 
        \bar{P}^{(31)}_{2MSW}(\nu_e \rightarrow \nu_e ;t_{\odot},t_0).~~~\eqno(19)$$
\noindent The three-neutrino mixing solar
$\nu_e$ survival probability in the case of interest is always smaller than 
the corresponding two-neutrino MSW probability. 

  Eqs. (18) and (19) determine the range in which
the probability $\bar{P}(\nu_e \rightarrow \nu_e ;t_E,t_0)$ can oscillate
as a function of $R/p$ due to the presence of the vacuum oscillation component 
$P^{(21)}_{2VO}$ in it. Let us consider several specific cases relevant to our further
discussion.

  If the solar $\nu_e$ undergo extreme nonadiabatic MSW transitions in the 
Sun ($p/\Delta m^2_{31}$ is ``large''), one has \cite{SP200} 
$P'_{(31)} \cong \cos^2\theta_{13}$, 
$\bar{P}^{(31)}_{2MSW} \cong 1 - {1\over {2}}\sin^22\theta_{13}$ and 
$$|U_{e3}|^4 = \sin^4\theta_{13} \leq \bar{P}(\nu_e \rightarrow \nu_e ; t_E,t_0) 
\leq |U_{e3}|^4 +~(1 - |U_{e3}|^2)^2 = 
\sin^4\theta_{13} + \cos^4\theta_{13}.~~\eqno(20)$$
\noindent We get the same result, eq. (20), 
if $N_e(t_0)(1 - \tan2\theta_{13})^{-1} < N_e^{res}$, i.e.,
when $p/\Delta m^2_{31}$ is sufficiently small. 
In this case \cite{KP88} the $\nu_e$ MSW transitions 
are adiabatic ($P'_{(31)} \cong 0$),
the $\nu_e$ oscillate in the Sun due to $\Delta m^2_{31}$ as in vacuum: 
$\cos2\theta_{13}^{m}(t_0) \cong \cos2\theta_{13}$ and 
$\bar{P}^{(31)}_{2MSW} \cong 1 - {1\over {2}}\sin^22\theta_{13}$.
If, however, in the ``small'' $p/\Delta m^2_{31}$ region the vacuum oscillation term
$\cos 2\pi R/L^{v}_{21}$ averages out so that effectively  
$P^{(21)}_{2VO} = 1 - 1/2~\sin^22\theta_{12}$, one obtains: 
$$\bar{P}(\nu_e \rightarrow \nu_e ; t_E,t_0) \cong |U_{e1}|^4 + |U_{e2}|^4 +
 |U_{e3}|^4 = 1 - {1\over {2}}\sin^22\theta_{12}~\cos^4\theta_{13} - 
      {1\over {2}}\sin^22\theta_{13},~~~~\eqno(21)$$
\noindent which is the average three-neutrino vacuum oscillation probability.

  If the solar $\nu_e$ MSW transitions are 
adiabatic (i.e., $P'_{(31)} \cong 0$) and
$\cos2\theta_{13}^{m}(t_0) \cong -1$ 
(i.e., $N_e(t_0)/N_e^{res} \gg 1, \tan2\theta_{13}$, solar neutrinos are born
``above'' and ``far'' (in $N_e$) from the resonance region), one has
$\bar{P}^{(31)}_{2MSW} \cong |U_{e3}|^2$ and therefore
$$\bar{P}(\nu_e \rightarrow \nu_e ; t_E,t_0) \cong 
   \bar{P}^{(31)}_{2MSW} \cong |U_{e3}|^2 = \sin^2\theta_{13}.~~~\eqno(22)$$
\noindent Under the above conditions the $\nu_e$ state in matter at the point of 
$\nu_e$ production (in the Sun) essentially coincides with the heaviest
of the three neutrino matter-eigenstates, which continuously evolves 
(as the neutrino propagates towards the surface of the Sun)
into the mass (energy) eigenstate $|\nu_3>$ 
at the surface of the Sun. As a consequence,
vacuum oscillations do not take place between the Sun and the Earth and 
$\bar{P}(\nu_e \rightarrow \nu_e ; t_E,t_0)$ coincides with the probability
to find $\nu_e$ in the state $|\nu_3>$.

  Let us discuss next the properties of the probability 
$\bar{P}(\nu_e \rightarrow \nu_e ; t_E,t_0)$ if inequality (5), i.e., a Zee model 
type relation between $\Delta m^2_{32}$ and $\Delta m^2_{31}$ 
($\Delta m^2_{21}$), is valid and $\Delta m^2_{32}$ is assumed to give rise
to solar neutrino long wave length vacuum oscillations on the path between
the Sun and the Earth. The probability 
$\bar{P}(\nu_e \rightarrow \nu_e ; t_E,t_0)$ in this case has the form:
 $$\bar{P}^{Z}(\nu_e \rightarrow \nu_e ; t_E,t_0) = 
 \bar{P}^{(13)}_{2MSW}(\nu_e \rightarrow \nu_e ;t_{\odot},t_0)~~~~~~~~~~~~~~~~~~~
~~~~~~~~~~~~~~~~~~~~~~~~~~~~~~~~~~~~~~~~~$$ 
$$~~~~~~~~~~~~~~~~~ +~~{{1 - |U_{e1}|^2}\over {1 - 2|U_{e1}|^2}}~[1 -  
  P^{(23)}_{2VO}(\nu_e \rightarrow \nu_e ;t_E,t_{\odot})]~
  [|U_{e1}|^2 - \bar{P}^{(13)}_{2MSW}(\nu_e \rightarrow \nu_e ;t_{\odot},t_0)]
         ~\eqno(23a)$$
$$~~~~~~~~ = \bar{P}^{(13)}_{2MSW} 
- {1\over {2}}~\sin^22\theta_{23}~(1 - \cos2\pi {R \over {L^{v}_{32}}})~
\sin^2\theta'_{13}[{1\over {2}} - ({1\over {2}} - P'_{(13)})
\cos2\theta_{13}^{'m}(t_0)].~~\eqno(23b)$$
\noindent The angles $\theta_{23}$ and $\theta'_{13}$ in eq. (23b) 
are determined by
$$\sin^22\theta_{23} = 4 {{|U_{e3}|^2~|U_{e2}|^2}\over {(|U_{e3}|^2 + 
|U_{e2}|^2)^2}}~~,~~~~\eqno(24)$$
and 
$$\cos2\theta'_{13} = 2|U_{e1}|^2 - 1 > 0,~~
 \sin2\theta'_{13} = - 2|U_{e1}|\sqrt{1 - 
|U_{e1}|^2}~ < 0,~~~~\eqno(25)$$
\noindent  and $L^{v}_{32} = 4\pi p/\Delta m^2_{32}$; the requirement 
$$|U_{e1}|^2 > 0.5~~~~\eqno(26)$$
\noindent guarantees the possibility of resonant enhancement 
of the MSW transitions in the Sun. The two-neutrino mixing MSW probability
$\bar{P}^{(13)}_{2MSW}(\nu_e \rightarrow \nu_e ;t_{\odot},t_0) \equiv
 \bar{P}^{(13)}_{2MSW}$ is given by eq. (10) in which the angles
$\theta_{13}$, $\theta_{13}^{m}(t_0)$ and the ``jump'' probability 
$P'_{(31)}$ are replaced respectively by $\theta'_{13}$, 
$\theta_{13}^{'m}(t_0)$ and $P'_{(13)}$, where 
$\theta_{13}^{'m}(t_0)$ and $P'_{(13)}$ are defined by eq. (13) and by eq. (15) 
in which $\theta_{13}$ is substituted by $\theta'_{13}$. The two-neutrino
vacuum oscillation probability 
$P^{(23)}_{2VO}(\nu_e \rightarrow \nu_e ;t_E,t_{\odot}) \equiv P^{(23)}_{2VO}$ 
can be obtained by replacing in eq. (11) $\theta_{12}$ and $L^{v}_{21}$ with
$\theta_{23}$ and $L^{v}_{32}$.

  Expressions (23) follow 
\footnote{In the particular case of the Zee model lepton mixing matrix $U$
\cite{Zee,SPZee} in which two of the three mixing angles have specific values
and all elements of $U$ are real, expression (23a) for 
$\bar{P}^{Z}(\nu_e \rightarrow \nu_e ; t_E,t_0)$ reduces to the one derived
in ref. \cite{ASTZee} in the indicated case.}
from eqs. (9) and (17) if we 
interchange the indices 1 and 3 in the quantities 
$|U_{ej}|^2$ and $\Delta m^2_{ik}$, or equivalently, if we replace
$|U_{e3(1)}|^2$ with $|U_{e1(3)}|^2$ and $\Delta m^2_{31}$
($\Delta m^2_{21}$) with $\Delta m^2_{13}$ ($\Delta m^2_{23}$) in all the terms
entering into eqs. (9) and (17), except in the ``jump'' probability
$P'_{(31)}$, and take into account the definitions of $\theta_{13}$ (eq. (12))
and $\theta'_{13}$ (eq. (24)). It can be shown that  
$P'_{(13)}$ coincides with the expression for $P'_{(31)}$, eq. (15),  in which 
$\theta_{13}$ is changed to $\theta'_{13}$.

Although expressions (9) ((17)) and (23a) ((23b)) formally look quite similar,
the two probabilities $\bar{P}(\nu_e \rightarrow \nu_e ; t_E,t_0)$ and
$\bar{P}^{Z}(\nu_e \rightarrow \nu_e ; t_E,t_0)$ actually have 
very different properties.

  Expression (17) implies that if 
$\Delta m^2_{21} \ll \Delta m^2_{31}$, a genuinely hybrid small mixing angle,
$\sin^2\theta_{13} \ll 1$, MSW transition + vacuum oscillation solution(s)
of the solar neutrino problem is (are) in principle possible. In contrast,
if $\sin^2\theta'_{13} \ll 1$ the contribution of the vacuum oscillation probability
$P^{(23)}_{2VO}$ in $\bar{P}^{Z}(\nu_e \rightarrow \nu_e ; t_E,t_0)$
is strongly suppressed and $\bar{P}^{Z}(\nu_e \rightarrow \nu_e ; t_E,t_0) \cong
\bar{P}^{(13)}_{2MSW}(\nu_e \rightarrow \nu_e ;t_{\odot},t_0)$. Thus, 
if inequality (5) holds, hybrid MSW + VO solutions
can exist only for large values of the MSW mixing angle $\theta'_{13}$.

  Similarly to $\bar{P}(\nu_e \rightarrow \nu_e ; t_E,t_0)$, the probability
$\bar{P}^{Z}(\nu_e \rightarrow \nu_e ; t_E,t_0)$
is limited from above by $\bar{P}^{(13)}_{2MSW}$:
$\bar{P}^{Z}(\nu_e \rightarrow \nu_e ; t_E,t_0) \leq
\bar{P}^{(13)}_{2MSW}(\nu_e \rightarrow \nu_e ;t_{\odot},t_0)$
\footnote{For the specific case of the Zee model mixing matrix $U$
\cite{Zee,SPZee} this upper bound was obtained in ref. \cite{ASTZee}.}. 
This follows from
the fact that $0.5 < |U_{e1}|^2 \leq 1.0$ and 
$[|U_{e1}|^2 - \bar{P}^{(13)}_{2MSW}] > 0$
\footnote{We have used the upper bound 
$\bar{P}^{(13)}_{2MSW} \leq 1 - 2|U_{e1}|^2(1 - |U_{e1}|^2) = 
1 - 1/2~\sin^22\theta'_{13}$ to get the last inequality.}. 
However, the lower limit
on  $\bar{P}^{Z}(\nu_e \rightarrow \nu_e ; t_E,t_0)$ is physically very different
from the lower limit on $\bar{P}(\nu_e \rightarrow \nu_e ; t_E,t_0)$, eq. (18):
$$\bar{P}^{Z}(\nu_e \rightarrow \nu_e ;t_E,t_0) \geq
{{|U_{e1}|^2}\over {2|U_{e1}|^2 - 1}}~[\bar{P}^{(13)}_{2MSW} - (1 - |U_{e1}|^2)] =
\cos^2\theta'_{13} [{1\over {2}} + ({1\over {2}} - P'_{(13)})
\cos2\theta_{13}^{'m}(t_0)]~\eqno(27)$$

In the two ``asymptotic'' regions of  
``small'' (adiabatic MSW transitions) and ``large'' (extreme nonadiabatic 
MSW transitions) 
values of $p/\Delta m^2_{31}$ the probability  
$\bar{P}^{Z}(\nu_e \rightarrow \nu_e ;t_E,t_0)$
has the form:
$$\bar{P}^{Z}(\nu_e \rightarrow \nu_e ; t_E,t_0) \cong |U_{e1}|^4 +~
(1 - |U_{e1}|^2)^2 P^{(23)}_{2VO}(\nu_e \rightarrow \nu_e ;t_E,t_{\odot}).~~
                                    \eqno(28)$$
\noindent From eq. (28) we get the following 
band of allowed values of 
$\bar{P}^{Z}(\nu_e \rightarrow \nu_e ;t_E,t_0)$, 
which is the same for the two regions:
$$|U_{e1}|^4 = \cos^4\theta'_{13} \leq \bar{P}^{Z}(\nu_e \rightarrow 
                  \nu_e ; t_E,t_0) 
\leq |U_{e1}|^4 +~(1 - |U_{e1}|^2)^2 = 
\cos^4\theta'_{13} + \sin^4\theta'_{13}~.~~\eqno(29)$$
\noindent The corresponding band
\footnote{Note that the band of allowed values of
 $\bar{P}^{Z}(\nu_e \rightarrow \nu_e ;t_E,t_0)$ in the 
``large'' $p/\Delta m^2_{31}$ region
shown in Fig. 2 in ref. \cite{ASTZee} does not correspond to eqs. (28) and (29).}
for the probability 
$\bar{P}(\nu_e \rightarrow \nu_e ;t_E,t_0)$ is given in (20).

  Finally, if the MSW transitions in the Sun are adiabatic, $P'_{(13)} = 0$,
and  $\cos2\theta_{13}^{'m}(t_0) = - 1$  
($\bar{P}^{(13)}_{2MSW} = \sin^2\theta'_{13})$, one has:
$$0 \leq \bar{P}^{Z}(\nu_e \rightarrow \nu_e ;t_E,t_0) =
(1 - |U_{e1}|^2)P^{(23)}_{2VO}(\nu_e \rightarrow \nu_e ;t_E,t_{\odot}) 
\leq \sin^2\theta'_{13}~, \eqno(30)$$
\noindent to be compared with eq. (22). 

   It follows from the above discussion that if
$\Delta m^2_{21} \ll \Delta m^2_{31}$, 
the amplitude of the
oscillations of $\bar{P}(\nu_e \rightarrow \nu_e ; t_E,t_0)$ 
(due to the $P^{(21)}_{2VO}$ term) as $R/p$ varies 
is equal in the regions of ``large'' and 
``small'' $p/\Delta m^2_{31}$ to $\cos^4\theta_{13}\sin^22\theta_{12}$
and can be maximal; for $\sin^22\theta_{12} \sim 1$ it is always greater than 1/4. 
At the same time, this amplitude is strongly suppressed and vacuum oscillations
practically do not take place (even if $\sin^22\theta_{12} \sim 1$) in the
region of the adiabatic MSW transitions where $\cos2\theta_{13}^{m}(t_0) \cong -1$
and  $\bar{P}^{(31)}_{2MSW}$ has its minimal value,
$\bar{P}^{(31)}_{2MSW} = \sin^2\theta_{13}$. In contrast, if the Zee model type 
relation $\Delta m^2_{32} \ll \Delta m^2_{31}$ holds, the  
vacuum oscillations 
are suppressed in the two
``asymptotic'' regions: the amplitude of the oscillations is 
equal to $\sin^4\theta'_{13}\sin^22\theta_{23}$ and is 
always constrained to be smaller than 1/4. 
In the region of adiabatic MSW transitions,
where $\cos2\theta_{13}^{'m}(t_0) \cong -1$
and  $\bar{P}^{(13)}_{2MSW} = \sin^2\theta'_{13}$,
the probability $\bar{P}^{Z}(\nu_e \rightarrow \nu_e ; t_E,t_0)$
oscillates (due to $P^{(23)}_{2VO}$) as a function of $R/p$ 
with an amplitude $\sin^2\theta'_{13}\sin^22\theta_{23}$,
which does not exceed $0.5\sin^22\theta_{23}$. Obviously,
the vacuum oscillations can be important in this case
only for sufficiently large values of $\sin^2\theta'_{13}$.
 
  Because of the aforementioned differences in the properties of the probabilities 
$\bar{P}(\nu_e \rightarrow \nu_e ; t_E,t_0)$ and
$\bar{P}^{Z}(\nu_e \rightarrow \nu_e ; t_E,t_0)$, the case (4) of relations between
the neutrino mass squared differences provides
a much richer spectrum of possibilities of genuinely new
solutions of the solar neutrino problem of the hybrid MSW transition + vacuum 
oscillation type, than the case of relations (5). 

\vglue 0.4cm
\leftline{\bf 3. Hybrid MSW Transition + Vacuum Oscillation Solutions} 
\hskip 0.3truecm {\bf of the Solar Neutrino Problem }
\vskip 0.3cm
\indent We have analyzed the most recently published data from the
four solar neutrino experiments \cite{CHLOR,KAM,GALLEX,SAGE} searching for solutions 
of the solar neutrino 
problem of the hybrid MSW transitions + vacuum oscillations type. Only the 
case of inequality (4) with $\Delta m^2_{31}$ and $\Delta m^2_{21}$ having values
in the intervals (7) and (8) was studied. 

  The analysis we performed is based 
on the analytic expression eq. (9) for the average solar \nue survival probability,
$\bar{P}(\nu_e \rightarrow \nu_e ; t_E,t_0)$, with 
$\bar{P}^{(31)}_{2MSW}(\nu_e \rightarrow \nu_e ; t_{\odot},t_0)$ given by
eqs. (10), (13) and (15). As is well known, 
the solar neutrino flux consists of several components, 
six of which are relevant for the interpretation
of the results of the solar neutrino experiments \cite{SNP}
(see also, e.g., \cite{SNP1,SNP2}): the $pp$, $pep$, 
$^{7}$Be, $^{8}$B and the $CNO$ (two components). We utilized 
the predictions of the solar model of Bahcall and Pinsonneault from 1995 
with heavy element diffusion \cite{BP95} for the $pp$, $pep$, etc. neutrino fluxes 
in this study. The estimated uncertainties in the theoretical predictions for the 
indicated fluxes \cite{BP95} were not taken into account.
For each given neutrino flux the probability 
$\bar{P}(\nu_e \rightarrow \nu_e ; t_E,t_0)$ was averaged
over the corresponding region of \nue production in the Sun. Since the data
analyzed was accumulated over a period of several years, 
$\bar{P}(\nu_e \rightarrow \nu_e ; t_E,t_0)$ was also averaged over an interval of
time equal to one year \cite{KP1}. The probability 
$\bar{P}(\nu_e \rightarrow \nu_e ; t_E,t_0)$ depends on the time of the year $t_{y}$
through the dependence of $P^{21}_{2VO}(\nu_e \rightarrow \nu_e ; t_E,t_{\odot})$ 
on the distance between the Sun and the Earth, $t_E - t_{\odot} \cong R$,
which is a function of $t_{y}$:
$$R = R(t_{y}) = R_0~[1 - \epsilon \cos2\pi {t_{y}\over{T}}],~~~\eqno(31)$$
\noindent where $\epsilon = 0.0167$ is the ellipticity of the Earth 
orbit around the Sun and $T = 365~$ days. The solar neutrino fluxes are 
functions of $R^{-2}(t_{y})$ and their dependence on $t_{y}$ was 
effectively accounted for when we performed the time averaging of
the probability $\bar{P}(\nu_e \rightarrow \nu_e ; t_E,t_0)$.   
 
  The results we shall report have been obtained utilizing the $\chi^2-$ method. 
The regions of the relevant MSW transition and vacuum oscillation 
parameters allowed by the data were
determined by requiring that $\chi^2 < 3.841$. 
This choice is motivated by the following arguments. 
In the $\chi^2-$analysis in the case
under study there are zero degrees of freedom: we have four experimental results
and the theory tested contains four parameters ($\theta_{12}$,  
$\Delta m^2_{21}$, $\theta_{13}$ and $\Delta m^2_{31}$). 
Nevertheless, the existence of 
regions of values of the parameters for which the $\chi^2-$function 
has a distinct minimum and a sufficiently
low value at the minimum is a strong and unambiguous indication 
that the mechanism 
of solar \nue flux depletion considered gives a good quality 
of the fit of the data and therefore provides a solution of 
the solar neutrino problem. At the same time,
in the case of zero degrees of freedom it is impossible to assign a precise
confidence level to the $\chi^2 = 3.841$ contours. It should be noted,
however, that in grand unified theories with massive neutrinos the quantities
$\Delta m^2_{21}$  and $\Delta m^2_{31}$ are typically not independent: their
ratio is a function of the ratio of quark or charged lepton masses
\footnote{In certain classes of theories some of the lepton mixing angles
are also functions of the charged lepton or quark masses.}. If such a 
relation holds there would be three independent parameters in the theory, 
and correspondingly one degree of freedom in the $\chi^2-$analysis.
Under these conditions $\chi^2 = 3.841$ corresponds to 95\% C.L.

 We have searched for MSW + VO solutions 
of the solar neutrino problem
for values of $\sin^22\theta_{12}$ and $\sin^22\theta_{13}$ from the 
intervals:
$$ 0.20 \ltap \sin^22\theta_{12} \leq 1.0,~~~\eqno(32)$$
$$ 10^{-4} \leq \sin^22\theta_{13} < 1.0.~~~\eqno(33)$$    
\noindent The intervals (32) and (33) are considerably wider than 
the corresponding ones in the case of purely two-neutrino mixing
($2\nu$) vacuum oscillation and MSW transition solutions 
(see, e.g., \cite{KP2,MSW1,MSW2,KPnu96}).
We did not perform a complete scan of the 4-parameter space defined 
by eqs. (7), (8), (32) and (33). We were primarily interested 
in solutions which do not converge continuously (via regions where $\chi^2 \leq 3.841$)
to the $2\nu$ vacuum oscillation or MSW transition solution
in the limit of $\sin^22\theta_{13} \rightarrow 0$ or, correspondingly,
of $\sin^22\theta_{12} \rightarrow 0$ or 
$\Delta m^2_{12} \rightarrow 0$. Nevertheless, we have found five 
different regions of values of the four parameters, for which 
$\chi^2 \leq 3.841$. These regions (solutions) are denoted 
by $A$, $B$, $C$, $D$ and $E$
below. Sections of the indicated regions in the planes of 
the MSW transition and of the vacuum oscillation parameters 
$\Delta m^2_{31} - \sin^22\theta_{13}$ and 
$\Delta m^2_{21} - \sin^22\theta_{12}$ are shown respectively in
Figs. 1a, 1b and 2a, 2b. Shown in Figs. 1a -- 2b are also the points in which
the $\chi^2-$function has a minimal value. 

 We shall discuss next the physical features of the solutions found. A detailed
analysis of the implications of the MSW transition + vacuum oscillation solutions 
for the solar neutrino experiments Super-Kamiokande, SNO, ICARUS, BOREXINO, HELLAZ
and HERON will be given elsewhere. Here we will limit ourself only to 
qualitative remarks. 

 {\bf Solution A.} For this solution (see Figs. 1a and 2a) the vacuum oscillation
parameters lie in the intervals 
$$4\times 10^{-12}~eV^2 \ltap \Delta m^2_{21} 
         \ltap 8\times 10^{-12}~eV^2,~~~\eqno(34a)$$  
$$ 0.65 \ltap \sin^22\theta_{12} \leq 1.0,~~~\eqno(34b)$$
\noindent while the region formed by the 
values of the MSW transition parameters extends
from the domain of the $2\nu$ nonadiabatic solution at 
$\Delta m^2_{31} \cong 5.0\times 10^{-6}~eV^2$,  
$\sin^22\theta_{13} \cong 8\times 10^{-3}$, first to larger (smaller) values of 
$\Delta m^2_{31}$ ($\sin^22\theta_{13}$) reaching   
$\Delta m^2_{31} \cong 10^{-4}~eV^2$ and 
$\sin^22\theta_{13} \cong 3.1\times 10^{-4}$, and then to larger values of 
$\sin^22\theta_{13}$ up to 0.5, spanning approximately 3 orders of magnitude
in $\sin^22\theta_{13}$ at practically constant 
$\Delta m^2_{31} \cong (1.1 - 1.3) \times 10^{-4}~eV^2$.
The minimum $\chi^2$ value for this solution is 
$\chi^2_{min} \cong 0.071$ and corresponds 
to ($\Delta m^2_{21}$, $\sin^22\theta_{12}$, $\Delta m^2_{31}$, 
$\sin^22\theta_{13}$) 
$\cong$ ($5.6\times 10^{-12}~eV^2$, 0.98, $4.2\times 10^{-5}~eV^2$, $10^{-3}$) 
(the black triangle-down
in Fig. 1a). In the ``horizontal'' region where 
$\sin^22\theta_{13} \gtap  10^{-3}$  and 
$\Delta m^2_{31} \cong (1.1 - 1.3) \times 10^{-4}~eV^2$ there exists a 
local minimum with a somewhat larger value of
$\chi^2_{min}$: one finds $\chi^2_{min} \cong 1.0$.
This minimum is reached for 
($\Delta m^2_{21}$, $\sin^22\theta_{12}$, $\Delta m^2_{31}$, $\sin^22\theta_{13}$) 
$\cong$ ($5.4\times 10^{-12}~eV^2$, 1.0, $1.2\times 10^{-4}~eV^2$, 0.14) 
(the black dot in Fig. 1a).  

   For $p \cong E \geq 5~MeV$ and $\Delta m^2_{21}\leq 8.0\times 10^{-12}~eV^2$
we have $L^{v}_{21} \geq 1.6\times 10^{9}~km$, 
$\cos (2\pi R/L^{v}_{21}) \geq 0.82$ and, as follows from
eq. (11), $P^{(21)}_{2VO}(\nu_e \rightarrow \nu_e ; t_E,t_{\odot}) \geq 0.91$.
Thus, for $^{8}$B neutrinos with energy $E \geq 5~MeV$ one has
in the case of solution A:
$\bar{P}(\nu_e \rightarrow \nu_e ; t_E,t_0) \cong
\bar{P}^{(31)}_{2MSW}(\nu_e \rightarrow \nu_e ; t_{\odot},t_0)$, i.e.,
most of the $^{8}$B neutrinos undergo only MSW transitions
\footnote{Obviously, if, for instance,  
 $\Delta m^2_{21} = 4.0\times 10^{-12}~eV^2$, the same result will be valid
for the neutrinos with $E \geq 2.5~MeV$.}. 
The MSW transitions of the $^{8}$B neutrinos having energy
$E \geq 5~MeV$ are adiabatic for values of 
$\Delta m^2_{31} \cong (1.1 - 1.3) \times 10^{-4}~eV^2$
and $\sin^22\theta_{13}\gtap (3.0 - 4.0)\times 10^{-3}$ \cite{KP88}
from the ``horizontal'' region of the solution (see Fig. 1a).
They are nonadiabatic for values of 
$\Delta m^2_{31}$ and $\sin^22\theta_{13}$ 
from the remaining part of the allowed region.

   For the $pp$ and the major part of the $^{7}$Be neutrinos we have 
$E \leq 0.41~MeV$ and
$E = 0.862~MeV$, respectively, and for     
$\Delta m^2_{31} \gtap 1.3\times 10^{-5}~eV^2$,  
their energies fall in the ``small''
$p/\Delta m^2_{31}$ domain (see Section 2) where
$$\bar{P}(\nu_e \rightarrow \nu_e ; t_E,t_0) \cong
\sin^4\theta_{13} + 
\cos^4\theta_{13}~
P^{(21)}_{2VO}(\nu_e \rightarrow \nu_e ; t_E,t_{\odot}),~~\eqno(35)$$ 
\noindent with  $\sin^2\theta_{13} \ltap 0.15$ for the solution under discussion.
Thus, in contrast to the main fraction of $^{8}$B neutrinos, 
the $pp$ and $^{7}$Be \nue do not undergo resonant MSW transitions but 
take part in vacuum oscillations between the Sun and the Earth. 
Actually, the $^{7}$Be neutrino energy 
of $0.862~MeV$ is in the region of the first minimum of 
$P^{(21)}_{2VO}$ as $E$ decreases from the ``asymptotic'' values at which 
$P^{(21)}_{2VO} \cong 1$, while the interval of energies of the $pp$ neutrinos,
relevant for the current Ga--Ge and the presently discussed 
future solar neutrino experiments (HELLAZ, HERON), 
$0.22~MeV \ltap E \leq 0.41~MeV$, is in the region of the first 
maximum of $P^{(21)}_{2VO}$ as $E$ decreases further. The dependence of
the probability $\bar{P}(\nu_e \rightarrow \nu_e ; t_E,t_0)$ on the neutrino
energy E in the case of solution A is illustrated in Figs. 3a - 3c.

  For $5.0\times 10^{-6}~eV^2 \leq \Delta m^2_{31} \ltap 1.3 \times 10^{-5}~eV^2$, 
i.e., in the minor sub-region of solution A, which overlaps with the region of
the $2\nu$ nonadiabatic solution (see Fig. 1a), the 0.862 MeV $^{7}$Be 
neutrinos take part either in MSW transitions and vacuum oscillations or
in MSW transitions only. This small sub-region converges continuously to the
region of the $2\nu$ nonadiabatic solution when
$\sin^2\theta_{12} \rightarrow 0$ and we are not going to discuss it further.
Note that the domain of solution A in the $\Delta m^2_{21} - 
\sin^2\theta_{12}$ plane shown in Fig. 2a corresponds to values of 
$\Delta m^2_{31}$ and $\sin^2\theta_{13}$ which lie outside the sub-region
in question.  

   For solution A, the Kamiokande signal and the signal due to the $pp$ ($^{8}$B)
neutrinos in the Ga--Ge (Cl--Ar) experiments are smaller approximately by factors 
of $\sim (0.33 - 0.42)$ and $\sim (0.70 - 0.80)$ ($\sim 0.30 $), 
respectively, than the 
corresponding signals predicted in ref. \cite{BP95}. The $^{7}$Be $\nu_e$ 
flux is suppressed rather strongly -- by a factor of $\sim (0.13 - 0.14)$, 
with respect to the flux in the solar model of ref. \cite{BP95}.
   
  Let us note that the solution A region in the 
$\Delta m^2_{21} - \sin^2\theta_{12}$ plane of the vacuum oscillation
parameters is quite similar to the region
of the ``low'' $^{8}$B \nue flux 
$2\nu$ vacuum oscillation solution found in ref. \cite{KP1}. The latter is 
possible for values of the $^{8}$B neutrino flux which are lower by a factor
of 0.35 to 0.43 (of 0.30 to 0.37) than the flux predicted in \cite{BP92}
(in \cite{BP95} and used in the present study). Note, however, that the
$\chi^2_{min}$ for the indicated purely vacuum oscillation solution,
$\chi^2_{min} = 4.4~ (2~d.f.)$ \cite{KP1}, is
considerably larger than the value of $\chi^2_{min}$ for solution A. 
The two solutions differ drastically in the way the $^{8}$B 
neutrino flux is affected by the transitions and/or the oscillations.

  The implications of solution A for the future solar neutrino experiments
aimed at detection and studies of the $^{7}$Be and/or $pp$
neutrinos like BOREXINO, HELLAZ or HERON, are practically the same
as those discussed in detail in ref. \cite{KP1} for the case of the corresponding
purely vacuum oscillation ``low'' $^{8}$B neutrino flux solution. However,
the implications for the Super-Kamiokande, SNO and ICARUS experiments which
are sensitive only to $^{8}$B neutrinos are completely different. They will 
be discussed elsewhere. Let us mention only here that in the case of solution
A: i) the spectrum of $^{8}$B neutrinos will be strongly 
deformed (see Figs. 3a - 3c), 
ii) the magnitude of the day-night asymmetry 
in the signals in the indicated detectors 
can be very different from that predicted 
in the case of the $2\nu$ MSW solution (see, e.g., \cite{DN}), 
and iii) the seasonal variation of
the $^{8}$B $\nu_e$ flux \cite{SMBP78,KP2} practically coincides with the 
standard (geometrical)
one of 6.68\%.  

 {\bf Solution B.} For solution B (see Figs. 1a, 2a and 2b) 
the allowed region in the
$\Delta m^2_{31} - \sin^22\theta_{13}$ plane of the MSW transition parameters,
$$5.1\times 10^{-6}~eV^2 \ltap \Delta m^2_{31} 
         \ltap 1.2\times 10^{-5}~eV^2,~~~\eqno(36a)$$  
$$3.2\times 10^{-3} \ltap \sin^22\theta_{13} \ltap 6.6\times 10^{-3},~~~\eqno(36b)$$
\noindent looks approximately like the region of the $2\nu$
MSW nonadiabatic solution shifted as a whole to somewhat larger values of
$\Delta m^2_{31}$ (by a factor $\sim 1.3$) and to smaller values of
$\sin^22\theta_{13}$ (on average by a factor $\sim 1.6$). The two regions 
have a small common sub-region (see Fig. 1a) and our further discussion does not
extend to this sub-region of overlapping. Solution B can be regarded as
an ``improved'' MSW transitions + vacuum oscillations version of the
purely $2\nu$ MSW nonadiabatic solution.    

  In the $\Delta m^2_{21} - \sin^22\theta_{12}$ plane the region of solution B
extends in $\Delta m^2_{21}$ 
from $1.8\times 10^{-11}~eV^2$ at least up to 
$ 10^{-9}~eV^2$ (we did not perform a search for allowed values of
$\Delta m^2_{21}$ beyond $10^{-9}~eV^2$) and from 0.15 to practically
1.0 in $\sin^22\theta_{12}$:
$$0.15 \ltap \sin^22\theta_{12} \ltap 0.98.~~~\eqno(37)$$
\noindent For $\Delta m^2_{21}$ from the domain 
$\sim (0.5 - 1.0)\times 10^{-10}~eV^2$ of the
$2\nu$ vacuum oscillation solution \cite{KP1,KPnu96}, 
however, solution B takes place for values of $\sin^22\theta_{12}$ which
are systematically smaller than the values of the same parameter
in the $2\nu$ vacuum oscillation solution. As is seen in Fig. 2b, in the
$\Delta m^2_{21} - \sin^22\theta_{12}$ plane there are two marginally 
disconnected regions, which we will call ``down'' and ``up'':
the separation ``line'' is approximately at   
$\Delta m^2_{21} \cong 1.5\times 10^{-10}~eV^2$. Except in the region where
$\Delta m^2_{21} \cong (1.0 - 2.0)\times 10^{-10}~eV^2$, the vacuum 
oscillation mixing parameter is relatively small for solution B: one has
$\sin^22\theta_{12} \ltap 0.8$, and for most of the allowed values of
$\Delta m^2_{21}$ actually $\sin^22\theta_{12} \ltap 0.7$.

  The minimum $\chi^2$ value for the solution under discussion is even
smaller than for solution A: we have $\chi^2_{min} \cong 0.025$ at 
($\Delta m^2_{21}$, $\sin^22\theta_{12}$, $\Delta m^2_{31}$, $\sin^22\theta_{13}$) 
$\cong$ ($1.4\times 10^{-10}~eV^2$, 0.43, $9.9\times 10^{-6}~eV^2$, 
$4.0\times 10^{-3}$) (the black triangle-up in Figs. 1a and 2a) 
\footnote{It is very close to the lowest $\chi^2$ value possible, $\chi^2_{min} = 0.024$, 
with the set of the data \cite{CHLOR,KAM,GALLEX,SAGE} we use in the present analysis.}.

In the ``up'' region
in the $\Delta m^2_{21} - \sin^22\theta_{12}$ plane one has
$\chi^2_{min} \cong 0.03$ and this minimum is located at 
($\Delta m^2_{21}$, $\sin^22\theta_{12}$, $\Delta m^2_{31}$, $\sin^22\theta_{13}$) 
$\cong$ ($5.8\times 10^{-10}~eV^2$, 0.43, $ 9.9\times 10^{-6}~eV^2$, 
$3.8\times 10^{-3}$).

  For $E \leq 0.41~MeV$ and values of $\Delta m^2_{31}$ 
and $\sin^22\theta_{13}$ from the solution
intervals (36a) and (36b), one has 
$\bar{P}^{(31)}_{2MSW} \cong 1 - 1/2\sin^22\theta_{13} \cong 1$, and consequently
we obtain from eq. (9): $\bar{P}(\nu_e \rightarrow \nu_e ; t_E,t_0) \cong
P^{(21)}_{2VO}(\nu_e \rightarrow \nu_e ; t_E,t_{\odot})$. This implies that 
the effect of the MSW transitions for the $pp$ neutrinos is negligible and they
effectively take part in vacuum oscillations only. Moreover, for
$E \leq 0.41~MeV$ and $\Delta m^2_{21} \gtap 3\times 10^{-11}~eV^2$ we find
$L^{v}_{21} \ltap 3.4\times 10^{7}~km$ and $2\pi R/L^{v}_{21} \gtap 27$.
Correspondingly, for $E \leq 0.41~MeV$ the term $\cos (2\pi R/L^{v}_{21})$ in   
$P^{(21)}_{2VO}$ is a fastly oscillating function of $R/p$, which is averaged 
practically to zero, e.g., by the integration over the $pp$ neutrino energy, or
by averaging over an uncertainty $\Delta E$ in 
the measured value of the energy $E$,
$\Delta E/E \gg 4\times 10^{-2}$. Therefore for 
$\Delta m^2_{21} \gtap 3\times 10^{-11}~eV^2$
the $pp$ neutrino contribution
to the event rate of the Ga--Ge detectors, for instance, is suppressed  
by the energy-independent factor 
$\bar{P}(\nu_e \rightarrow \nu_e ; t_E,t_0) \cong
\bar{P}^{(21)}_{2VO} \cong 1 - 1/2 \sin^22\theta_{12}$.

  The 0.862 MeV $^{7}$Be neutrinos take part in adiabatic MSW transitions
in the Sun, while the $^{8}$B neutrinos with $E \gtap 4~MeV$ undergo
nonadiabatic transitions. Both the $^{7}$Be and $^{8}$B neutrinos,
as well as the $\nu_{\mu}$ and/or $\nu_{\tau}$ into which a fraction of the $\nu_e$
has been converted by the MSW effect in the Sun, participate in vacuum oscillations
after leaving the Sun. These oscillations are modulated by the MSW probability
$\bar{P}^{(31)}_{2MSW}$ (see Figs. 4a - 4b). Let us note that for 
$\Delta m^2_{21} \ltap 3.0\times 10^{-10}~eV^2$ the effects of the averaging 
of $\bar{P}(\nu_e \rightarrow \nu_e ; t_E,t_0)$ over the time period
of 1 year is negligible for the 0.862 MeV $^{7}$Be neutrinos; this effect becomes
important for the $^{7}$Be $\nu_e$ for 
$\Delta m^2_{21} \gtap 5.0\times 10^{-10}~eV^2$, as Fig. 4b illustrates.
For the range of $\Delta m^2_{21}$ of interest, 
$\Delta m^2_{21} \ltap (5.0 - 6.0)\times 10^{-10}~eV^2$,
the time averaging does not change the probability
$\bar{P}(\nu_e \rightarrow \nu_e ; t_E,t_0)$ for solar neutrino energies
$E \gtap 3.0~MeV$, i.e., for the dominant fraction of the $^{8}$B neutrinos.
    
    With respect to the predictions in ref. \cite{BP95}, the signal in the 
Kamiokande detector and the contribution of the $^{8}$B 
neutrinos to the signals in the Cl--Ar
detector are smaller typically by factors of
$\sim (0.43 - 0.47)$ and $\sim (0.32 - 0.36)$. 
The $pp$ and the 0.862 MeV $^{7}$Be $\nu_e$ fluxes
are suppressed by factors of $\sim (0.65 - 0.90)$ and $\sim (0.11 - 0.27)$ for 
most of the values of the parameters from the allowed region. 
However, for $\sin^22\theta_{12} \sim 0.9$, for instance, one has 
$\bar{P}(\nu_e \rightarrow \nu_e ; t_E,t_0) \cong 0.55$ 
for the $pp$ neutrinos. Even in this case the 0.862 MeV $^{7}$Be $\nu_e$ flux 
is reduced by a factor of $\sim 0.3$, but the indicated possibility is rather
marginal.  

    There are rather large and distinctive distortions of 
the $^{8}$B $\nu_e$ spectrum
(see Figs. 4a and 4b) in the case of solution B.  The seasonal variations 
due to the vacuum oscillations 
of the signals in the Super-Kamiokande, SNO and ICARUS detectors are 
estimated to be smaller than the variations in the case of the $2\nu$ vacuum
oscillation solution \footnote{These variations were shown \cite{KP2} to be
not larger than 15\% for the $2\nu$ vacuum oscillation solution.},
except possibly in the small region
of the  $\Delta m^2_{21} - \sin^22\theta_{12}$ plane where 
$\Delta m^2_{21} \gtap 10^{-10}~eV^2$ and $\sin^22\theta_{12} \gtap 0.7$. 
The range of the predicted values
of the day-night asymmetry in these detectors is different from
the one expected for the $2\nu$ MSW solution.
The seasonal variation of the 0.862 MeV $^{7}$Be $\nu_e$ flux caused 
by the vacuum oscillations is expected to be considerably smaller than in the 
$2\nu$ case, while the day-night asymmetry is estimated to be somewhat 
smaller than the one predicted for the $2\nu$ MSW nonadiabatic solution.
The seasonal variation, nevertheless, may be observable.
Obviously, the experimental detection 
both of a deviation from the standard (geometrical) 6.68\% seasonal variation 
of the solar neutrino flux and of a nonzero day-night effect will be a proof 
that solar neutrinos take part in MSW transitions and vacuum oscillations.
   
{\bf Solution C}. The values of the parameters corresponding to solution C 
(see Figs. 1a and 2a) form a rather large region in the 
$\Delta m^2_{31} - \sin^22\theta_{13}$ plane,
$$1.1\times 10^{-5}~eV^2 \ltap \Delta m^2_{13}
         \ltap 1.2\times 10^{-4}~eV^2,~~~\eqno(38a)$$  
$$1.7\times 10^{-4} \ltap \sin^22\theta_{13} \ltap 2.0\times 10^{-3},~~~\eqno(38b)$$
\noindent and a relatively small one in the 
$\Delta m^2_{21} - \sin^22\theta_{12}$ plane,
$$1.1\times 10^{-10}~eV^2 \ltap \Delta m^2_{21} 
         \ltap 1.3\times 10^{-10}~eV^2,~~~\eqno(39a)$$  
$$0.7\ltap \sin^22\theta_{12} \leq 1.0.~~~\eqno(39b)$$
  The $\chi^2_{min}$ for this solution is larger than for solutions A and B:
$\chi^2_{min} \cong 1.49$ at 
($\Delta m^2_{21}$, $\sin^22\theta_{12}$, $\Delta m^2_{31}$, $\sin^22\theta_{13}$) 
$\cong$ ($1.2\times 10^{-10}~eV^2$, 0.78, $4.6\times 10^{-5}~eV^2$, 
$5.9\times 10^{-4}$) (the black square in Figs. 1a and 2a). 

  For the values (38a) and (38b) of $\Delta m^2_{31}$ and $\sin^22\theta_{13}$
and the energies of the $pp$ neutrinos of interest, we have
$\bar{P}^{(31)}_{2MSW} \cong 1 - 1/2\sin^22\theta_{13} \cong 1$ and the $pp$
neutrino flux is effectively suppressed by the energy-independent factor
$\bar{P}(\nu_e \rightarrow \nu_e ; t_E,t_0) \cong
\bar{P}^{(21)}_{2VO} \cong 1 - 1/2\sin^22\theta_{12}$ (Figs. 5a and 5b). 
The energy of the dominant fraction of $^{7}$Be neutrinos, 
$0.862~MeV$, is for this solution in the region
of energies for which the effect of the time-averaging of the probability
$\bar{P}(\nu_e \rightarrow \nu_e ; t_E,t_0)$ is important. This effect is
negligible for $E \gtap 2~MeV$, i.e., for most of the $^{8}$B neutrinos. The 
0.862 MeV $^{7}$Be $\nu_e$ flux is suppressed either due to 
the vacuum oscillations only (Fig. 5b) or due to the combined effect of
the vacuum oscillations and MSW transitions (Fig. 5a). The suppression of the
$^{8}$B $\nu_e$ flux is a result of the interplay  
of the vacuum oscillations and the MSW transitions (Figs. 5a and 5b). 
As a consequence, i) the $^{8}$B $\nu_e$ spectrum
is strongly deformed and ii) one expects observably large seasonal variations due
to the vacuum oscillations of the $^{8}$B and/or, for
$\Delta m^2_{31} \gtap 3\times 10^{-5}~eV^2$, of the $^{7}$Be $\nu_e$ fluxes. 
The day-night asymmetry in the signals of the Super-Kamiokande, SNO and ICARUS
experiments is not expected to be larger than a few percent. In the case of
the $^{7}$Be $\nu_e$ flux the asymmetry is either absent 
(if $\Delta m^2_{31} \gtap 3\times 10^{-5}~eV^2$) or does not 
exceed a few percent. 
     
   In comparison with the predictions of ref. \cite{BP95}, 
the Kamiokande signal and the signals
produced by the $^{8}$B neutrinos in the  
Cl--Ar detector are reduced respectively by factors of
$\sim (0.35 - 0.41)$  and  
$\sim (0.24 - 0.32)$, while the ranges
of suppression of the $pp$ and of the 0.862 MeV $^{7}$Be $\nu_e$ 
fluxes read respectively 0.50 - 0.65 and 0.17 - 0.30.  

   {\bf Solution D.} In the $\Delta m^2_{31} - \sin^22\theta_{13}$ plane 
the region of solution D (see Figs. 1b and 2a) can formally be obtained
from the one of solution A by moving the latter as a whole
to larger values of $\Delta m^2_{31}$ and smaller values of  
$\sin^22\theta_{13}$ and by truncating two relatively small sub-regions
located at $\sin^22\theta_{13} \gtap 0.2$ and at 
$\Delta m^2_{31} \ltap 1.1\times 10^{-5}~eV^2$. In the 
$\Delta m^2_{21} - \sin^22\theta_{12}$ plane the solution D region is determined by
$$3.2\times 10^{-11}~eV^2 \ltap \Delta m^2_{21}
         \ltap 4.0\times 10^{-11}~eV^2,~~~\eqno(40a)$$  
$$0.62 \ltap \sin^22\theta_{12} \leq 1.0.~~~\eqno(40b)$$

   The minimum $\chi^2$ value for this solution is similar to that for solution
C, $\chi^2_{min} \cong 1.35$; it is located at 
($\Delta m^2_{21}$, $\sin^22\theta_{12}$, $\Delta m^2_{31}$, $\sin^22\theta_{13}$) 
$\cong$ ($3.4\times 10^{-11}~eV^2$, 0.80, $1.5\times 10^{-5}~eV^2$, 
$2.5\times 10^{-3}$) (the black diamond in Figs. 1b and 2a). 

  For the values of $\Delta m^2_{21}$ given in eq. (40a) and 
$0.22~MeV \leq E \leq 0.41~MeV$, the effect of the time averaging of the probability
$\bar{P}(\nu_e \rightarrow \nu_e ; t_E,t_0)$ cannot be neglected; for 
$E \gtap 0.7~MeV$ it becomes inessential. The $\nu_e$ survival probability for
the 0.862 MeV $^{7}$Be neutrinos is given by expression (35). The same expression is
approximately valid for the $pp$ neutrinos (see Figs. 6a - 6c). However,
the $pp$ neutrino contribution to the signals 
in the Ga-Ge detectors is suppressed by the factor given in eq. (21).
\noindent The $^{8}$B $\nu_e$ flux is suppressed due to the combined effect
of the MSW transitions and vacuum oscillations (Figs. 6a - 6c). 

   With respect to the predictions in ref. \cite{BP95}, the contributions of the
$pp$ and $^{7}$Be ($^{8}$B) neutrinos in the signals of the Ga--Ge
(Cl--Ar) detectors are smaller by factors of $\sim (0.50 - 0.70)$ and 
$\sim (0.31 - 0.49)$  ($\sim (0.22 - 0.28)$), 
while the signal in the Kamiokande detector
is reduced by a factor of $\sim (0.32 - 040)$.
   
   As in the case of solutions B and C, the $^{8}$B $\nu_e$ spectrum is 
strongly deformed by the combined effect of the MSW transitions and 
the vacuum oscillations. The seasonal variations of the $pp$ and of the 
0.862 MeV $^{7}$Be 
$\nu_e$ fluxes generated by the vacuum oscillations are estimated to be greater
than $\sim~$10\%; they are expected to be smaller than a few percent for
the $^{8}$B $\nu_e$  having energy $E \geq 5~MeV$. The day-night asymmetry
in the signals of the Super-Kamiokande, SNO and ICARUS detectors may be observable
only for values of $\sin^22\theta_{13}\gtap 10^{-3}$; even in this region
it is not expected to be bigger than a few percent. The day-night effect for
the 0.862 MeV $^{7}$Be $\nu_e$ flux is negligible.

  Solution D has as a $\theta_{13}\rightarrow 0$ limit the second
``low'' $^{8}$B $\nu_e$ flux $2\nu$ vacuum oscillation solution discussed in
ref. \cite{KP1}: the regions of values of the vacuum oscillation parameters
of the two solutions practically coincide. This $2\nu$ solution  
was found to be possible for values of the initial   
$^{8}$B $\nu_e$ flux which are by a factor $\sim (0.45 - 0.65)$ 
($\sim (0.39 - 0.56)$) smaller than
the flux predicted in ref. \cite{BP92} (in ref. \cite{BP95}). The value 
of $\chi^2_{min}$ for the indicated
purely $2\nu$ vacuum oscillation solution, 
$\chi^2_{min} \cong 5.0$ (2 d.f.) \cite{KP1},
is considerably larger than $\chi^2_{min}$ for the solution D.
The MSW + VO effects and correspondingly the purely VO effects on
the $^{7}$Be and/or $^{8}$B neutrino fluxes in the cases of 
the two solutions are also very different.

  {\bf Solution E.} This solution holds for relatively large
values of $\sin^22\theta_{12}$ and $\sin^22\theta_{13}$, 
$$0.70 \ltap \sin^22\theta_{12} \ltap 0.80,~~~\eqno(41a)$$
$$0.14 \ltap \sin^22\theta_{13} \ltap 0.39,~~~\eqno(41b)$$
\noindent and values of $\Delta m^2_{21}$ forming a tiny region around the point
$\Delta m^2_{21} \cong 9.4\times 10^{-10}~eV^2$, with the values of  
$\Delta m^2_{31}$ belonging to the narrow interval
$1.2\times 10^{-4}~eV^2 \ltap \Delta m^2_{31}
         \ltap 1.4\times 10^{-4}~eV^2$ (Figs. 1b and 2b). The $\chi^2_{min}$ value
is rather large, $\chi^2_{min} \cong 3.3$, and we shall not discuss this solution in
greater detail. Let us mention only that the $pp$ and $^{7}$Be  $\nu_e$ fluxes 
are suppressed by one and the same energy-independent 
factor given in eq. (21) - the average
$\nu_e$ survival probability in the case of $3\nu$ vacuum oscillations,
while both MSW transitions and vacuum oscillations contribute to the
suppression of the $^{8}$B $\nu_e$ flux. This is illustrated in Fig. 7.  

  Finally, we would like to point out that, as Fig. 8 indicates, 
the MSW transitions + vacuum oscillations may provide an alternative 
to the purely MSW mechanism of suppression of the
solar $\nu_e$ flux in the case discussed in ref. \cite{BeLWPK}.

  \vglue 0.4cm
\leftline{\bf 4. Conclusions.} 
\vskip 0.3cm
    Assuming three flavour neutrino mixing takes place in vacuum, we have
investigated the possibility that the observed depletion of the solar $\nu_e$
flux is caused by an interplay of MSW transitions of the solar $\nu_e$ in the Sun
followed by long wave length neutrino oscillations in vacuum 
on the way from the Sun to the Earth.
Choosing the neutrino mass squared difference $\Delta m^2_{31} > 0$ to 
be responsible for the MSW transitions in the Sun, 
the long wave length vacuum oscillations can be due either to
$\Delta m^2_{21} > 0$ (inequality (4) is valid) or to $\Delta m^2_{32} > 0$ 
((5) holds). In both cases the solar $\nu_e$ survival probability is
described, as we have shown, by simple analytic 
expressions, eq. (9) and eq. (23a). 
However, the two cases were found to be very different in their
physics implications. In particular, if (5) is realized,  
the vacuum oscillations are not suppressed, in principle, 
only for large values of the MSW mixing angle parameter.
The case of (4) appeared to provide a much richer spectrum of possible
new solutions of the solar neutrino problem of the hybrid 
MSW transitions + vacuum oscillations type and we have analyzed only this case.   
We have found for the values of $\Delta m^2_{21}$ and $\Delta m^2_{31}$ 
within the ranges given in (7) and (8) several versions of this solution,
denoted by A, B, C, D, and E in the text, most of which 
are distinctly different from the two-neutrino 
mixing purely vacuum oscillation or MSW solution.
 In particular, the values of the MSW parameters $\Delta m^2_{31}$ and
$\sin^22\theta_{13}$ of the different solutions found lie outside the 
regions of the $2\nu$ MSW solutions
(including the regions obtained by varying the $^{8}$B and $^{7}$Be neutrino fluxes)
\cite{KP2,MSW1,MSW2,KPnu96}, except in two cases 
(solutions A and B) in which very small sub-regions 
are located within the region of the $2\nu$ MSW nonadiabatic solution.
Two of the solutions studied, A and D, 
are similar in what regards the effective mechanism and the magnitude 
of suppression of the $pp$ flux
to the two ``low'' $^{8}$B $\nu_e$ flux $2\nu$ vacuum oscillation
solutions discussed in \cite{KP1}. However, the $\chi^2_{min}$ values for the former
are much smaller than for the latter. In addition, the MSW + VO solutions
A and D and the two corresponding purely vacuum oscillation 
solutions \cite{KP1} imply very different mechanisms of 
suppression of the $^{7}$Be and/or the $^{8}$B neutrino fluxes.  
  
    A general feature of the MSW + VO 
solutions studied by us is that
the $pp~~\nu_e$ flux is suppressed (albeit not strongly - 
by a factor not smaller than 0.5) 
primarily due to the 
vacuum oscillations of the $\nu_e$, the suppression of the 
0.862 MeV $^{7}$Be $\nu_e$ flux 
is caused either by the vacuum oscillations or by the combined effect of the 
MSW transitions and the vacuum oscillations, 
while the $^{8}$B $\nu_e$ flux is suppressed either due to the
MSW transitions only or by the interplay of the MSW transitions 
in the Sun and the oscillations in vacuum on the way to the Earth. 
The solutions differ in the way the $pp$, $^{7}$Be and 
the $^{8}$B neutrinos are affected by the $\nu_e$
MSW transitions and/or the oscillations in vacuum.

  Searching for MSW + VO solutions we did not scan
the entire region of the parameter space defined by eqs. (7), (8), (32) and (33).
In particular, solutions which correspond to a suppression (by a factor $\gtap 0.5$)
of the $pp$ neutrino flux due to the MSW transitions, concomitant
with the requisite suppression of the $^{7}$Be and of the $^{8}$B neutrino fluxes
due to the MSW effect and/or the vacuum oscillations and respectively the
vacuum oscillations alone, are possible for 
$\Delta m^2_{31} \sim (10^{-8} - 10^{-7})~eV^2$
and $\Delta m^2_{21} \sim (0.3 - 1.0)~10^{-10}~eV^2$.
 
  For all MSW + VO solutions we have 
considered, the $^{8}$B $\nu_e$ spectrum is predicted to be rather strongly 
deformed. The $^{8}$B neutrinos undergo nonadiabatic transitions for 
values of the MSW parameters 
$\Delta m^2_{31} \cong (0.1 - 1.2) \times 10^{-4}~eV^2$ and
$\sin^22\theta_{13} \cong (0.2 - 3.0)\times 10^{-3}$
which are respectively larger and smaller than the values
corresponding to the $2\nu$ MSW nonadiabatic solution. The $^{8}$B $\nu_e$
adiabatic transitions take place for 
$\Delta m^2_{31} \cong (1.1 - 1.5)\times 10^{-4}~eV^2$ and
$\sin^22\theta_{13} \cong (3\times 10^{-3} - 0.5)$. Clearly, the Kamiokande data 
on the shape of the $^{8}$B neutrino spectrum can be used to further constrain the
solutions we have found. Such an analysis, however, lies outside the scope
of the present study. 

   For the MSW + VO solutions considered by us
the day-night asymmetry in the signals of the detectors sensitive only to
$^{8}$B or $^{7}$Be neutrinos are estimated to be rather small, not exceeding
a few percent. The seasonal variation effect caused by the vacuum oscillations
can be observable for $^{7}$Be neutrinos and, for certain 
relatively small regions of the allowed values of the parameters, 
can also be observable for the $^{8}$B or for 
the $pp$ neutrinos if the $pp$  neutrino flux 
is measured with detectors like HELLAZ or HERON.  
 
  Finally, the MSW transitions + vacuum oscillations can be an alternative
to the purely MSW mechanism of suppression of the solar \nue flux in the case
considered in ref. \cite{BeLWPK}.

\vglue 0.4cm
\leftline{\bf Acknowledgements.} We would like to thank Y. Suzuki for sending us
information about the detection efficiency of the Kamiokande detectors.
S.T.P. wishes to thank K. Babu, C.N. Leung, 
E. Lisi and L. Wolfenstein for useful discussions. 
Q.Y.L. would like to acknowledge helpful discussions with A.Yu. Smirnov. 
The work of S.T.P. was supported in part 
by the EEC grant ERBFMRXCT960090 and by Grant PH-510 from the
Bulgarian Science Foundation. 
\vskip 0.3cm

\vglue 0.4cm
\leftline{\bf Appendix. Derivation of the Expression for the 
       Solar \nue Survival Probability}
\vskip 0.3cm
\indent In this Appendix we present a derivation of expression
(9) for the average probability $\bar{P}(\nu_e \rightarrow \nu_e ; t_E,t_0)$,
which is not based on the general analytic result for the average solar \nue
survival probability in the case of three-neutrino mixing 
$\bar{P}(\nu_e \rightarrow \nu_e ; t,t_0)$, derived in ref. \cite{3nuSP}.
We remind the reader that the 
two relevant MSW transition and vacuum oscillation
parameters $\Delta m^2_{31}$ and $\Delta m^2_{21}$ (or $\Delta m^2_{32}$)
are assumed to satisfy inequality (4) (or (5)) and to lie in the intervals 
(8) and (7), respectively.  

  The probability amplitude that the solar 
\nue will not be converted into \numu and/or \nutau 
when it propagates from the central part of the Sun, 
where it was produced at time $t_0$, to the surface of 
the Earth, reached at time $t_E$, $A(\nu_e \rightarrow \nu_e ; t_E , t_0)$, 
can be written in the following form: 
 $$A(\nu_e \rightarrow \nu_e ; t_E , t_{0}) = \sum_{ l=e,\mu,\tau}
A_{MSW}(\nu_e \rightarrow \nu_l ; t_{\odot} , t_0)  
A_{VO}(\nu_l \rightarrow \nu_e ; t_E , t_{\odot} ), ~~\eqno(A1)$$
Here $t_\odot$ is the time at which the neutrino 
reaches the surface of the Sun, 
$A_{MSW}(\nu_e \rightarrow \nu_l ; t_{\odot} , t_0)$ is the amplitude of the
probability to find neutrino $\nu_l$, $l=e,\mu,\tau$, 
at the surface of the Sun 
after the $\nu_e$ underwent MSW 
transitions into $\nu_{\mu}$ and/or $\nu_{\tau}$
while propagating in the Sun, and 
$A_{VO}(\nu_l \rightarrow \nu_e ; t_E , t_{\odot})$ 
is the amplitude of the probability that $\nu_l$ will oscillate into
$\nu_e$ while traveling 
from the surface of the Sun to the surface of the Earth. 
For stable and relativistic neutrinos $\nu_k$ the amplitude 
$A_{VO}(\nu_l \rightarrow \nu_e ; t_E , t_{\odot})$ is given by the well-known
expression (see, e.g., \cite{SMBP78,BPet87}):
$$A_{VO}(\nu_l \rightarrow \nu_e ; t_E , t_{\odot}) = \sum_ {k=1}^3
U_{ek} e^{-iE_{k}(t_E - t_{\odot})}U_{lk}^* \cong 
e^{-iE_{1}R}\sum_ {k=1}^3
U_{ek} e^{-i{\Delta m^2_{k1}\over{2p}}R}U_{lk}^*, ~~\eqno(A2)$$
\noindent where $R = (t_E - t_{\odot}) \cong R_{0} = 1.4966\times 10^{8}~km$,
$R_{0}$ being the mean Sun-Earth distance, 
and we have used $E_{k} - E_{1} \cong \Delta m^2_{k1}/2p$.

 Consider first the case specified by inequality (4), with $\Delta m^2_{31}$ 
(and $\Delta m^2_{32}$) having values in 
the interval (7). For $p \leq 14.4~MeV$,
which is the maximal energy of the solar neutrinos detected on Earth,
one has $2p/\Delta m^2_{31(32)} \ltap 5.6\times 10^{4}~km$, and therefore
$(\Delta m^2_{31(32)}/2p)R_{0} \gtap 2.7\times 10^{3}$.
As a consequence the phases $(\Delta m^2_{31}/2p)R$ and
$(\Delta m^2_{32}/2p)R$ in eq. (A2) give rise to fastly oscillating
(with the change of $p$ and/or $R$) terms in the probability of solar
\nue survival:
$$P(\nu_e \rightarrow \nu_e ; t_E,t_0) = 
     |A(\nu_e \rightarrow \nu_e ; t_E , t_{0})|^2~.~~~~\eqno(A3)$$

\noindent When $P(\nu_e \rightarrow \nu_e ; t_E,t_0)$ is averaged
over (see, e.g., \cite{SMBP78,BPet87}) the region of neutrino production in the Sun, and/or
the uncertainty of the energy of the neutrino detected on Earth, and/or
over the uncertainty in the position of the solar neutrino detector
\footnote{This is relevant only for the radiochemical detectors (Homestake \cite{CHLOR},
GALLEX \cite{GALLEX} and SAGE \cite{SAGE}).}, and/or is integrated over a relevant solar 
neutrino energy interval, the fastly oscillating terms become 
strongly suppressed and their contribution in the averaged probability,
$\bar{P}(\nu_e \rightarrow \nu_e)$, can be
neglected. Taking this effectively into account by omitting the indicated 
fastly oscillating terms and using eqs. (A1) - (A3) one obtains an
expression for the average probability, 
$\bar{P}(\nu_e \rightarrow \nu_e ; t_E,t_0)$,
which is convenient to represent as a sum of two terms: 
$$\bar{P}(\nu_e \rightarrow \nu_e ; t_E,t_0) = \bar{S_1}(t_{\odot},t_0) + 
\bar{S_2}(t_E,t_{\odot},t_0) \equiv \bar{S_1} + \bar{S_2}~,~~~\eqno(A4)$$ 
\noindent where
$$S_1 = \sum_{k=1}^3~\sum_{l,l'=e,\mu,\tau} |U_{ek}|^2
(U_{lk}^* A_{MSW}(\nu_e \rightarrow \nu_l; t_{\odot},t_0))
(U_{l'k}^* A_{MSW}(\nu_e \rightarrow \nu_{l'}; t_{\odot},t_0))^*~,\eqno(A5)$$
$$S_2 = 2Re~[e^{-i{\Delta m^{2}_{21}\over{2p}}R}
 \sum_{l,l'=e,\mu,\tau} (U_{e2}
            A_{MSW}(\nu_e \rightarrow \nu_l; t_{\odot},t_0)U_{l2}^*)
 (U_{e1}A_{MSW}(\nu_e \rightarrow \nu_{l'}; t_{\odot},t_0)U_{l'1}^*)^*],
  ~\eqno(A6)$$ 
\noindent and the bar on the quantities $S_{1,2}$ in eq. (A4) means that all fastly
oscillating terms in $S_{1,2}$ arising from the amplitudes
$A_{MSW}(\nu_e \rightarrow \nu_l; t_{\odot},t_0)$, which are 
rendered negligible by the averaging
over the region of neutrino production in the Sun, etc. should be dropped.
Obviously, $\bar{S_1}(t_{\odot},t_0)$ is determined only by the solar \nue
transitions in the Sun, while $\bar{S_2}(t_E,t_{\odot},t_0)$ contains all the
dependence of $\bar{P}(\nu_e \rightarrow \nu_e ; t_E,t_0)$ on the long
wave length vacuum oscillations between the Sun and the Earth. Note, however,
that $\bar{S_2}(t_E,t_{\odot},t_0)$ depends on the 
\nue transitions in the Sun as well.  

  The MSW transition amplitudes $A_{MSW}(\nu_e \rightarrow \nu_l; t,t_0) \equiv
A_{l}(t)$, $t \leq t_{\odot}$, $l=e,\mu,\tau$, satisfy the system of evolution
equations:
$$i {{d} \over {dt}} A_l(t) = \sum_{l'=e,\mu ,\tau} M_{ll'} (t)A_{l'}(t). 
~\eqno(A7)$$
Here M(t) is the evolution matrix which can be 
chosen in the form (see, e.g., \cite{3nuMSW,3nuSP}):
$$M_{ll'}(t)={1\over {2p}}[\sum_{j=2}^3 U_{lj} \Delta m_{j1}^2 U_{jl'}^+ +
A(t)\delta _{el}  \delta _{ll'} - 
\delta_{ll'}(\sum _{k=2}^3 {|U_{ek}|}^2 \Delta m_{k1}^2 +
A(t))], ~~\eqno(A8)$$ 
where 
$$A(t) = 2p{\sqrt 2}G_FN_e(t), ~~\eqno(A9)$$
\noindent and $N_e(t)$ is the value of the electron number density at 
the point of the neutrino trajectory in the Sun, reached at time $t$. The
initial conditions for the system of equations (A8) read:
$$A_e(t_0) = 1, A_{\mu}(t_0) = A_{\tau}(t_0) = 0.~~~\eqno(A10)$$

  Since $\Delta m_{21}^2 \leq 5\times 10^{-10}~eV^2$ and 
$\Delta m_{31}^2 \geq 10^{-7}~eV^2$, we can expect that the evolution 
of the neutrino system in the Sun practically does not depend on
$\Delta m_{21}^2$. At the same time 
the lower energy pp neutrinos, for instance, 
can take part in vacuum oscillations in the Sun 
induced by $\Delta m_{21}^2$ \cite{SP200,SP88osc,SPJR89}. 
Inspecting eq. (A8) for the evolution 
matrix $M(t)$ one 
could conclude that the terms containing $\Delta m_{21}^2$
as a factor can be neglected provided 
$|U_{l3} \Delta m_{31}^2 U_{l'3}^*| \gg |U_{l2} \Delta m_{21}^2 U_{l'2}^*|$,
$l\neq l'=e,\mu,\tau$, and  
$|(|U_{l3}|^2 - |U_{e3}|^2)\Delta m_{31}^2|  \gg
|(|U_{l2}|^2 - |U_{e2}|^2)\Delta m_{21}^2|$, $l=e,\mu$. Actually, using some of the
intermediate results of the analysis performed in ref. \cite{3nuSP} one can
convince oneself that for $\Delta m_{21}^2$ and 
$\Delta m_{31}^2$ having values in the intervals (7) and (8) the effects
of $\Delta m_{21}^2$ on the solar neutrino transitions/oscillations in the Sun
are either effectively accounted for by the term $\bar{S}_{2}(t_E,t_{\odot},t_0)$ 
or are negligible (even for the pp neutrinos). 
This implies that one can formally set
$\Delta m_{21}^2$ to zero in the expression for $M(t)$, eq. (A8). In this case
the evolution matrix $M(t)$ depends only on the elements $U_{l3}$, 
$l=e,\mu,\tau$, forming the third column of the matrix $U$. 

  The phases of the elements 
$U_{l3}$, $U_{l3} = e^{i\alpha_{l3}}|U_{l3}|$, 
where $\alpha_{l3}$ are real constants,
can be eliminated from $M(t)$ by a redefinition of the amplitudes
$A_l(t)$: $A_l(t) \rightarrow e^{-i\alpha_{l3}}A_l(t)$, $l=e,\mu,\tau$. The 
evolution matrix $\bar{M}(t)$ of the system 
of evolution equations for the amplitudes
$e^{-i\alpha_{l3}}A_l(t)$ can formally be obtained from $M(t)$ by replacing
$U_{l3}$ by $|U_{l3}|$ in the expression (A8) for $M(t)$. As we shall show,
the average solar \nue survival probability 
$\bar{P}(\nu_e \rightarrow \nu_e ; t_E,t_0)$
does not depend in the case of interest on the 
phases of the elements of the lepton
mixing matrix $U$, and, in particular, on $\alpha_{l3}$.

 Performing the transformation 
$$\left({\begin{array}{c} e^{-i\alpha_{e3}}A_e(t)\\e^{-i\alpha_{\mu 3}}A_\mu(t)\\
e^{-i\alpha_{\tau 3}}A_\tau(t) \end{array}}\right) =
\left(\begin{array}{ccc} 1 & 0 & 0\\0 & |U_{\tau 3}|\over {\sqrt{
|U_{\tau 3}|^2 + |U_{\mu 3}|^2}} & |U_{\mu 3}|\over {\sqrt{|U_{\tau 3}|^2 +
 |U_{\mu 3}|^2}}\\0 & -{|U_{\mu 3}|
  \over {\sqrt{|U_{\tau 3}|^2 + |U_{\mu 3}|^2}}}
& |U_{\tau 3}|\over {\sqrt{|U_{\tau 3}|^2 + |U_{\mu 3}|^2}} \end{array} \right)
\left({\begin{array}{c} e^{-i\alpha_{e3}}A_e(t) \\A'_{\mu'}(t) \\
e^{-i\alpha_{e3}}A'_{\tau'}(t) \end{array}}\right), 
~\eqno(A11)$$ 

\noindent we find that the amplitudes 
($A_e(t)$,$A'_{\mu'}(t)$,$A'_{\tau'}(t)$) 
satisfy a system
of evolution equations having the same form as the system of equations (A7)
with an evolution matrix $M'(t)$ given by the expression:
$$M'(t) = {1 \over{2p}} \left[{\begin{array}{ccc} 0 & 0 & |U_{e3}|\sqrt{1 - 
          |U_{e3}|^2}~ 
\Delta m_{31}^2 \\ 0 & - |U_{e3}|^2 \Delta m_{31}^2 - A(t) & 0 \\ 
|U_{e3}|\sqrt{1 - |U_{e3}|^2}~\Delta m_{31}^2 & 0 & (1-2|U_{e3}|^2)
\Delta m_{31}^2-A(t)\end{array}}\right]~\eqno(A12)$$
\noindent In obtaining eq. (A12) we have used the unitarity 
of the lepton mixing matrix
$U$: $|U_{e3}|^2 + |U_{\mu 3}|^2 + |U_{\tau 3}|^2 = 1$. 
The initial conditions for the
new system of equations follow from eqs. (A10) and (A11):
$$A_e(t_0) = 1, A'_{\mu'}(t_0) = A'_{\tau'}(t_0) = 0.~~~\eqno(A13)$$

 It follows from eq. (A12) that the evolution of the amplitude 
$A'_{\mu'}(t)$ is decoupled from the evolution of 
$A_{e}(t)$ and  $A'_{\tau'}(t)$, while 
$A_{e}(t)$ and  $A'_{\tau'}(t)$ satisfy a 
two-neutrino mixing system of evolution
equations describing two-neutrino MSW 
transitions of the solar \nue in the Sun. 
The transitions can be enhanced by the solar matter effects,
as it follows from (A12), provided $|U_{e3}|^2 < 0.5$.
Explicit expressions for the amplitudes $A_{e}(t)$ and  $A'_{\tau'}(t)$ 
have been obtained, in particular, 
assuming $N_e(t)$ decreases exponentially along 
the neutrino path in the Sun \cite{SP200,SP88osc,AASP92}. In this case
the system of evolution equations for $A_{e}(t)$ and  $A'_{\tau'}(t)$ can be solved
exactly \cite{SP200,Kaneko}. Let us note that the exponential 
approximation describes with a good 
precision the change of $N_e(t)$ along the 
neutrino path, predicted by the solar models \cite{SNP,BP95}.

 As it is easy to see, eqs. (A11) - (A13) (see also (A8)) imply: 
$$A'_{\mu'}(t) = 0,~~~\eqno(A14a)$$
$$A_{\mu}(t) = {U_{\mu 3}\over {\sqrt{|U_{\mu 3}|^2 +
 |U_{\tau 3}|^2}}}e^{-i\alpha_{e3}}A'_{\tau'}(t),~~~\eqno(A14b)$$   
$$A_{\tau}(t) = {U_{\tau 3}\over {\sqrt{|U_{\tau 3}|^2 +
 |U_{\mu 3}|^2}}}e^{-i\alpha_{e3}}A'_{\tau'}(t).~~~\eqno(A14c)$$ 
\noindent Evidently, one has:
$$|A_e(t)|^2 + |A_{\mu}(t)|^2 + |A_{\tau}(t)|^2 = 
    |A_e(t)|^2 + |A'_{\tau'}(t)|^2 = 1.~~~\eqno(A15)$$

  We shall express next the probability $\bar{P}(\nu_e \rightarrow \nu_e ; t_E,t_0)$
in terms of the two amplitudes $A_{e}(t)$ and  $A'_{\tau'}(t)$ which
contain the whole information about the MSW transitions of neutrinos in the Sun.  
Inserting expressions (A14b) and (A14c) for $A_{\mu}(t)$ and
$A_{\tau}(t)$ in  equations (A5) and (A6) and using the unitarity of the lepton
mixing matrix U we obtain very simple 
expressions for the quantities $S_1$ and $S_2$:
$$S_1(t_{\odot},t_0) = [(1 - |U_{e3}|^2)^2 - 2|U_{e1}|^2|U_{e2}|^2]~
|A_{e}(t_{\odot}) - {|U_{e3}|\over {\sqrt{1 - |U_{e3}|^2}}}~
     A'_{\tau'}(t_{\odot})|^2$$
$$~~~~+ |U_{e3}|^2~|A_{e}(t_{\odot})|U_{e3}| + 
\sqrt{1 - |U_{e3}|^2}~A'_{\tau'}(t_{\odot})|^2,~~~\eqno(A16)$$
$$S_2(t_E,t_{\odot},t_0) = 2|U_{e1}|^2|U_{e2}|^2~|A_{e}(t_{\odot}) - 
{|U_{e3}|\over{\sqrt{1 - |U_{e3}|^2}}}~A'_{\tau'}(t_{\odot})|^2~
\cos2\pi {R\over{L^{v}_{21}}}.~~~\eqno(A17)$$ 

 Let us denote 
$$|A_e(t)|^2 = 1 - |A'_{\tau'}(t)|^2 \equiv 
P^{(31)}_{2MSW}(\nu_e \rightarrow \nu_e ;t_{\odot},t_0)~.~~~\eqno(A18)$$
\noindent Explicit expression for the probability 
$P^{(31)}_{2MSW}(\nu_e \rightarrow \nu_e ;t_{\odot},t_0)$
can be found in ref. \cite{SP88osc}. However, as we have already emphasized,
the oscillating terms in $P^{(31)}_{2MSW}(\nu_e \rightarrow \nu_e ;t_{\odot},t_0)$
give a negligible contribution in the averaged solar \nue survival probability
\cite{SPJR89}. Therefore only the average term 
$\bar{P}^{(31)}_{2MSW}(\nu_e \rightarrow \nu_e ;t_{\odot},t_0)$
in $P^{(31)}_{2MSW}(\nu_e \rightarrow \nu_e ;t_{\odot},t_0)$ should be taken
into account when one obtains $\bar{S}_1(t_{\odot},t_0)$ and
$\bar{S}_2(t_E,t_{\odot},t_0)$ from eqs. (A16) - (A18).

 The last quantity to be computed is the interference term 
${\rm Re}[A^{*}_e(t_{\odot})A'_{\tau'}(t_{\odot})]$ in eqs. (A16) and (A17).
This term coincides with the quantity ${\rm Re}[R_{H}(t'_0,t_0)]$ in eq. (29) in
ref. \cite{3nuSP} when $t'_0 = t_{\odot}$. An explicit analytic expression for 
${\rm Re}(A^{*}_e(t_{\odot})A'_{\tau'}(t_{\odot}))$ can be found exploiting, e.g.,
the exact analytic results for $A_e(t_{\odot})$ and $A'_{\tau'}(t_{\odot})$    
derived in ref. \cite{SP200} (see eqs. (28) and (29)) and cast in a form which
is more convenient to use in calculations like the present
one in ref. \cite{SP88osc,AASP92}. It follows from
eqs. (11), (12) and (18) - (20) in ref. \cite{AASP92}, for instance, that 
$${\rm Re}[A^{*}_e(t_{\odot})A'_{\tau'}(t_{\odot})] =   
-~({1\over{2}} - P'_{(31)})
  \sin2\theta_{13}\cos2\theta_{13}^{m}(t_0) + Oscillating~~terms,~~~~\eqno(A19)$$
\noindent where the angles $\theta_{13}$, $\theta_{13}^{m}(t_0)$ and the 
``jump'' probability $P'_{(31)}$ are defined in eqs. (12), (13) and (15).
The oscillating terms appearing in eq. (A19) do not contribute in the averaged
solar \nue survival probability 
\footnote{Explicit analytic expression for the oscillating terms
in eq. (A19) can be obtained from the formulae given in refs.
\cite{SP200,SP88osc,AASP92}.}. Let us note that, as can be shown \cite{SISSA45},
the result eq. (A19) for ${\rm Re}[A^{*}_e(t_{\odot})A'_{\tau'}(t_{\odot})]$ is general:
the fact that we have used the specific 
(but exact) exponential density solutions for 
$A_e(t_{\odot})$ and $A'_{\tau'}(t_{\odot})$ to derive it is reflected only
in the specific form of the ``jump'' probability $P'_{(31)}$  (eq. (15)) one gets
in eq. (A19).
    
 Taking into account the remarks we have made concerning the oscillating
terms in eqs. (A18) and (A19) and utilizing eqs. (A16) - (A19) and (10) - (14)
it is not difficult to get eq. (9) from eq. (A4).

  If the Zee model \cite{Zee} type relation (5) between 
$\Delta m^2_{21}$, $\Delta m^2_{31}$ and
$\Delta m^2_{32}$ is valid, we have to repeat the above analysis
interchanging the indices 1 and 3 in the quantities $U_{lj}$ and in 
$\Delta m^2_{ki}$ entering into eqs. (A6), (A8), (A11), (A12), (A14b),
(A14c), (A16) - (A18).
In particular, the evolution matrix $M'(t)$ in 
the system of evolution equations for the 
amplitudes ($A_e(t)$,$A'_{\mu'}(t)$,$A'_{\tau'}(t)$) now has the form:
$$M'(t) = {1 \over{2p}} \left[{\begin{array}{ccc} 0 & 0 & 
|U_{e1}|\sqrt{1 - |U_{e1}|^2}~ 
\Delta m_{13}^2 \\ 0 & - |U_{e1}|^2 \Delta m_{13}^2 - A(t) & 0 \\ 
|U_{e1}|\sqrt{1 - |U_{e1}|^2}~\Delta m_{13}^2 & 0 & (1-2|U_{e1}|^2)
\Delta m_{13}^2-A(t)\end{array}}\right]~,~\eqno(A20)$$
 
\noindent where $\Delta m_{13}^2 < 0$. The new form of $M'(t)$ and eqs. (A16) and 
(A17) in which $|U_{e3(1)}|^2$ and $L^{v}_{21}$ are now replaced respectively by 
$|U_{e1(3)}|^2$ and $L^{v}_{23}$, entail the introduction of the new angles   
$\theta_{23}$, $\theta'_{13}$ and  $\theta_{13}^{'m}(t_0)$ defined
by eqs. (24) - (26), as well as of the ``jump'' probability $P'_{(13)}$. 
It also leads to the requirement $|U_{e1}|^2 > 0.5$ which is a necessary
condition for the MSW resonance.  
Comparing the evolution matrices (A12) and (A20) it is easy to deduce
the form of $P'_{(13)}$ given the expression for $P'_{(31)}$, eq. (15).
The interference term ${\rm Re}[A^{*}_e(t_{\odot})A'_{\tau'}(t_{\odot})]$
is given by eq. (A19) in which $\theta_{13}$, $\theta_{13}^{m}(t_0)$ 
and $P'_{(31)}$ are replaced by $\theta'_{13}$, $\theta_{13}^{'m}(t_0)$ 
and $P'_{(13)}$. Using the above remarks it is easy to get expression
(24a) for the average probability 
$\bar{P}^{Z}(\nu_e \rightarrow \nu_e ;t_{E},t_0)$.

\newpage
\vskip 0.4cm
\centerline{\bf Figure Captions}
\medskip
\noindent
{\bf Figs. 1a -- 1b.} Regions of values of the
MSW transition parameters $\Delta m^2_{31}$
and $\sin^22\theta_{13}$ for which the $\chi^2 -$analysis of the
solar neutrino data gives $\chi^2 \leq 3.841$. The regions shown
are obtained for fixed values of the vacuum oscillation parameters
$\Delta m^2_{21}$ and $\sin^22\theta_{12}$ which are indicated in the figures.
Shown are also the 95\% C.L. $2\nu$ MSW nonadiabatic and 
adiabatic solution regions.

\medskip
\noindent
{\bf Figs. 2a -- 2b.} Regions of values of the
vacuum oscillation parameters $\Delta m^2_{21}$
and $\sin^22\theta_{12}$ allowed by the
solar neutrino data ($\chi^2 \leq 3.841$). The regions correspond to
fixed values of the MSW transition parameters
$\Delta m^2_{31}$ and $\sin^22\theta_{13}$ indicated in the figures.

\medskip
\noindent
{\bf Figs. 3a -- 3c.} The averaged probability 
$\bar{P}(\nu_{e} \rightarrow \nu_{e})~$ (solid line) as a function
of the solar neutrino energy $E$ for four sets of values of the
parameters indicated in the figures and belonging to region A. The 
dependence on $E$ of the 
MSW probability $\bar{P}^{(31)}_{2MSW}(\nu_{e} \rightarrow \nu_{e})~$
averaged over the region of the $^{8}$B neutrino production is also shown
(dashed line).

\medskip
\noindent
{\bf Figs. 4a -- 4b.} The probability 
$\bar{P}(\nu_{e} \rightarrow \nu_{e})~$ (solid line) not averaged (a) and 
averaged (b) over a period of 1 year \cite{KP1} (see the text), 
as a function of the neutrino energy
$E$ for two different sets of values of the four parameters from region B.
The averaged MSW probability $\bar{P}^{(31)}_{2MSW}(\nu_{e} \rightarrow \nu_{e})~$
as a function of $E$ is also shown (dashed line).

\medskip
\noindent
{\bf Figs. 5a -- 5b, 6a - 6c, 7.} Examples of the dependence of the averaged
probabilities $\bar{P}(\nu_{e} \rightarrow \nu_{e})~$ (solid line) and
$\bar{P}^{(31)}_{2MSW}(\nu_{e} \rightarrow \nu_{e})~$ (dashed line)
on the neutrino energy $E$ in the cases of solutions C, D and E,
respectively, considered in the text.

\medskip
\noindent
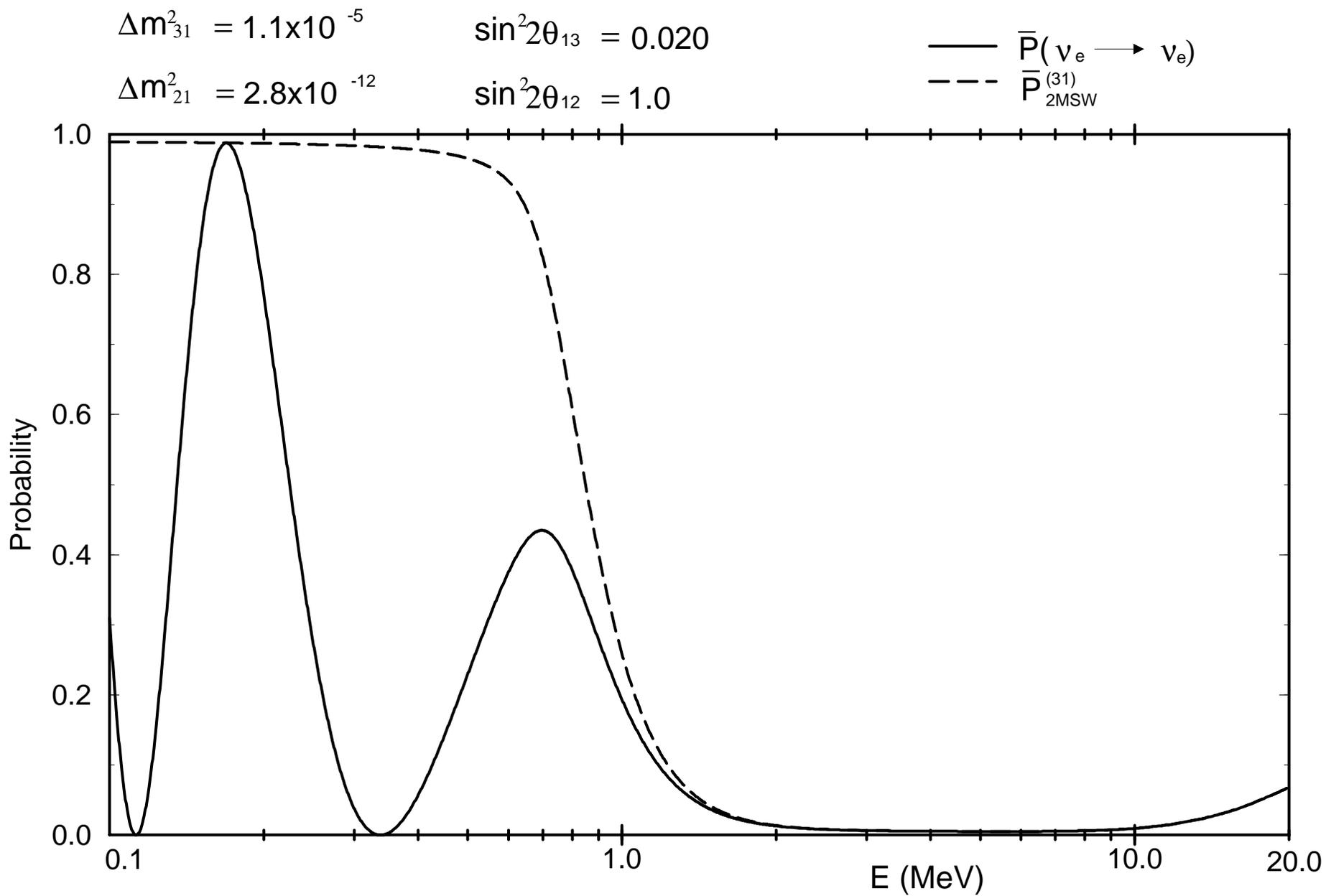
{\bf Fig. 8.} The averaged probability 
$\bar{P}(\nu_{e} \rightarrow \nu_{e})$ (solid line)
as a function of $E$ for values of the parameters (indicated in the figure),
for which the MSW transitions + vacuum oscillations can provide an alternative
to the purely MSW mechanism of suppression of the solar \nue flux
in the case discussed in ref. \cite{BeLWPK}.

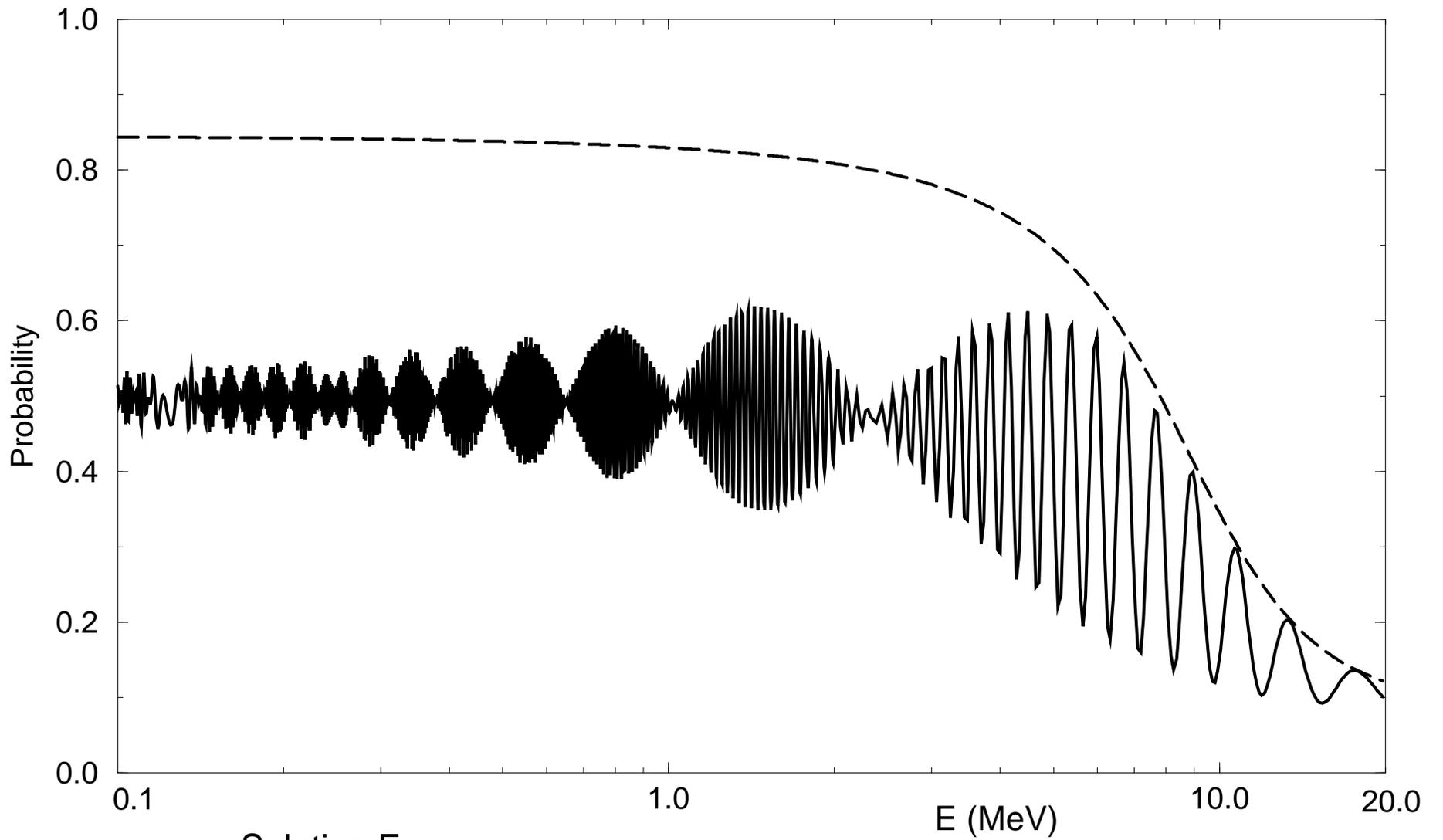
\end{document}